\documentclass[12pt]{article}
\usepackage{multirow}
\usepackage{graphics}
\usepackage{lscape}
\usepackage{amsfonts}
\usepackage{amsmath}
\usepackage{amssymb}
\usepackage{rotating}
\usepackage{amssymb,amsmath}
\usepackage[usenames]{color}
\usepackage{graphicx}
\usepackage[english]{babel}
\usepackage{natbib}
\usepackage{array}
\newcolumntype{P}[1]{>{\centering\arraybackslash}p{#1}}
\usepackage{array,multirow}
\usepackage{hhline}

\usepackage{float}
\usepackage{xcolor}
\usepackage{ulem} 

\usepackage{actuarialangle}
\usepackage[flushleft]{threeparttable}
\usepackage{comment}
\makeatletter

\newcommand{\Rmnum}[1]{\expandafter\@slowromancap\romannumeral #1@}
\makeatother
\usepackage{etoolbox}
\appto\TPTnoteSettings{\footnotesize}
\renewcommand{\section}[1]{\refstepcounter{section}\begin{center}{\sc \thesection. #1}\end{center}}
\renewcommand{\subsection}[1]{\bigskip \noindent \refstepcounter{subsection}{\thesubsection\ {\it #1}}}
\renewcommand{\subsubsection}[1]{\bigskip \noindent \refstepcounter{subsubsection}{\thesubsubsection\ {\it #1}}}
\renewcommand{\title}[1]{{\bf \begin{center}#1\end{center}}}
\renewcommand{\author}[1]{\begin{center}{\sc #1}\end{center}}

\newcommand{\acknowledgements}{\begin{center}{\sc Acknowledgements}\end{center}}
\newcommand{\competinginterests}{\begin{center}{\sc Competing Interests}\end{center}}
\renewenvironment{abstract}{\begin{center}{\sc abstract}\end{center}\small}{\normalsize}
\newenvironment{keywords}{\begin{center}{\sc keywords}\end{center}\small}{\normalsize}

\newenvironment{contacts}{\begin{center}{\sc contact addresses}\end{center}\small}{\normalsize}

\newenvironment{bajlist}{\begin{list}{(\alph{bajlistnum})\hfill}{\usecounter{bajlistnum}\setlength{\labelwidth}{0.3in}\setlength{\leftmargin}{0.3in}\setlength{\rightmargin}{0in}\setlength{\labelsep}{0in}\setlength{\topsep}{0in}\setlength{\partopsep}{0in}\setlength{\itemsep}{0in}\setlength{\parsep}{0in}}}{\end{list}{}\bigskip}
\newenvironment{bajsublist}{\begin{list}{(\arabic{bajsublistnum})\hfill}{\usecounter{bajsublistnum}\setlength{\labelwidth}{0.3in}\setlength{\leftmargin}{0.3in}\setlength{\rightmargin}{0in}\setlength{\labelsep}{0in}\setlength{\topsep}{0in}\setlength{\partopsep}{0in}\setlength{\itemsep}{0in}\setlength{\parsep}{0in}}}{\end{list}{}}

\newtheorem{proposition}{Proposition}
\newtheorem{corollary}{Corollary}

\def\ordd[#1][#2,#3;#4,#5]{\renewcommand{\arraystretch}{0.2}%
                            \setlength{\arraycolsep}{0pt}%
                            \begin{array}{ccc}%
                                 \scriptstyle#2 &  & \scriptstyle#3\\%
                                 \scriptstyle#4 &#1& \scriptstyle#5%
                           \end{array}}

\newcommand{\Prob}{\mbox{P}}
\newcommand{\Mean}{\mbox{E}}

\newcounter{bajlistnum}
\newcounter{bajsublistnum}

\begin{document}

\newcommand{\HRule}{\rule{\linewidth}{0.5mm}} 


\title{ON TECHNICAL BASES AND SURPLUS IN LIFE INSURANCE}

\author{By Oytun Ha\c{c}ar{\i}z\dag, Torsten Kleinow\ddag $\,$ and  Angus S. Macdonald\S}

\begin{abstract}

We revisit surplus on general life insurance contracts, represented by Markov models. We classify technical bases in terms of boundary conditions in Thiele's equation(s), allowing more general regulations than Scandinavian-style `first-order/second-order' regimes, and replacing the traditional retrospective policy value. We propose a `canonical' model with three technical bases (premium, valuation, accumulation) and show how each pair of bases defines premium loadings and surplus. Along with a `true' or `real-world' experience basis, this expands fundamental results of \cite{ramlau-hansen1988a}. We conclude with two applications: lapse-supported business; and the retrospectively-oriented regime proposed by \cite{moeller2007}.

\end{abstract}


\begin{contacts}

\noindent \dag $\,$ Department of Actuarial Sciences, The Faculty of Business, Karab{\"{u}}k University, Karab{\"{u}}k, 78050, Turkey, and Institute of Applied Mathematics, Middle East Technical University, Ankara, 06800, Turkey.

\noindent \ddag $\,$ Research Centre for Longevity Risk, Faculty of Economics and Business, University of Amsterdam.

\noindent \S $\,$ Department of Actuarial Mathematics and Statistics, Heriot-Watt University, Edinburgh EH14 4AS, UK, and the Maxwell Institute for Mathematical Sciences, UK.

\noindent Corresponding author: Angus Macdonald, A.S.Macdonald@hw.ac.uk.

\end{contacts}

\section{Introduction}
\label{sec:Intro}

\subsection{Motivation --- Unfinished Business}
\label{sec:Motivation}


Hoem introduced Markov models into life insurance mathematics \citep{hoem1969, hoem1988} and showed how life insurance cashflows could be represented by counting processes \citep{hoem1978}. These ideas were developed in several directions in the following years, see Section \ref{sec:IntroDevelopment}. We highlight two.

\begin{bajlist}

\item {\em Information}: The r\^ole of information, in terms of underlying $\sigma$-algebras {and filtrations}, was studied in \cite{norberg1991} and elsewhere. This helped to clarify the notion of retrospective and prospective policy values, as conditional expectations, see Section \ref{sec:StochDecomp}.

\item {\em Surplus}: Emerging surplus was studied in \cite{ramlau-hansen1988a}, \cite{linnemann1993} and others. In particular the first of these studies defined the surplus process as a stochastic object and decomposed it into an expected term plus a martingale error term\footnote{This idea has been taken up recently (see \cite{schilling2020}, \cite{jetses2022}) to define a similar decomposition of surplus into its contributions from each of several risk sources.} (Section \ref{sec:RamlauHansen1988}). 

\end{bajlist}

\noindent Almost in passing, the great classical results of life insurance mathematics were extended and clarified. \cite{hoem1969, hoem1988} had generalized Thiele's differential equation (henceforth just Thiele's equation) to Markov models. \cite{wolthuis1987}, \cite{ramlau-hansen1988b} and \cite{norberg1992} in particular showed Hattendorff's theorem \citep{hattendorff1868} to have been a prescient statement of a martingale property and Lidstone's theorem \citep{lidstone1905} was given modern form by \cite{norberg1985}. 

However, this entire literature took as given a Scandinavian style of insurance regulation, in which premiums and reserves were calculated on the same {\em first-order technical basis}. The experience was then {defined} by a {\em second-order technical basis} which, all being well, resulted in positive surpluses. See, for example, \cite{linnemann1993} and \cite{norberg1999}, {or for a discrete-time version \cite{olivieri2015}}. Requiring valuations to use the premium basis is more restrictive than in many jurisdictions {(see Section \ref{sec:Technical Bases})}; in particular it leaves two significant gaps.

\begin{bajlist}

\item It rules out common approaches such as net premium and gross premium valuations on bases chosen by the valuation actuary. 

\item The ability to vary the valuation basis, with {considerable} freedom, means that the actuary can choose when surplus will emerge, either sooner or later. This has obvious and important consequences. It is of interest to study how the choice of technical bases influences surplus, following in the footsteps of \cite{lidstone1905}.

\end{bajlist}

The example which motivated this work is the well-known result 
that {the expected present value (EPV) of} the total surplus earned in respect of a life insurance policy does not depend on the valuation basis used. {It implicitly assumes that premiums, valuations and the experience rest on separate technical bases.} The following, from a standard (and vintage) UK text on life insurance practice, is typical:

\begin{quote}
\small ``The effect of changing the valuation basis is to modify the [uniform] rate of emergence of profit $\ldots$.  The total real profit is unaffected, however, since it is a function of the premium and experience bases which cannot be affected by the valuation basis.'' \cite{fisher1965}.
\end{quote}


This proposition (when formulated more precisely) is just one of a suite of results about the incidence of surplus arising from relationships between {pairs of} technical bases. It is part of the third of the relationships introduced informally below labelled {\bf R1} to {\bf R3}.

\begin{itemize}
\item[{\bf R1}:] Initial surplus (strain) at inception of the policy arises from the difference between {\em premium} and {\em valuation} bases, and does not depend on the {\em experience} basis.
\item[{\bf R2}:] Surplus emerging during the policy term arises from the difference between the {\em valuation} and {\em experience} bases, and does not depend on the {\em premium} basis.
\item[{\bf R3}:] The EPV of total profit at the end of the term: (i) arises from the difference between the {\em premium} and {\em experience} bases; (ii) is equal to the EPV of the {total} surpluses in {\bf R1} and {\bf R2} and; (iii) does not depend on the {\em valuation} basis.
\end{itemize}


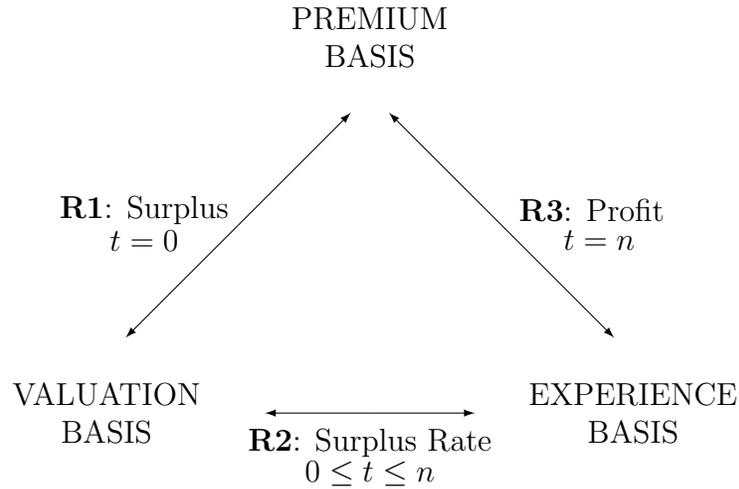
\begin{figure}
\begin{center}
\begin{picture}(100,65)
\put(50,60.5){\makebox(0,0)[c]{PREMIUM}}
\put(50,55.5){\makebox(0,0)[c]{BASIS}}
\put(15,10.5){\makebox(0,0)[c]{VALUATION}}
\put(15,5.5){\makebox(0,0)[c]{BASIS}}
\put(85,10.5){\makebox(0,0)[c]{EXPERIENCE}}
\put(85,5.5){\makebox(0,0)[c]{BASIS}}
\put(32.5,33){\vector(1,1){15}}
\put(32.5,33){\vector(-1,-1){15}}
\put(67.5,33){\vector(1,-1){15}}
\put(67.5,33){\vector(-1,1){15}}
\put(50,8){\vector(1,0){14}}
\put(50,8){\vector(-1,0){14}}
\put(20.0,35){\makebox(0,0)[c]{{\bf R1}: Surplus}}
\put(20.0,31){\makebox(0,0)[c]{$t = 0$}}
\put(79.5,35){\makebox(0,0)[c]{{\bf R3}: Profit}}
\put(80.5,31){\makebox(0,0)[c]{$t = n$}}
\put(50,4){\makebox(0,0)[c]{{\bf R2}: Surplus Rate}}
\put(50,0){\makebox(0,0)[c]{$0 \le t \le n$}}
\end{picture}
\end{center}
\caption{\label{fig:RelationsBases} Relationships between the premium basis, valuation (policy value) basis and {experience} basis.}
\end{figure}

\markright{On Technical Bases and Surplus in Life Insurance}


\noindent The relationships are illustrated in Figure \ref{fig:RelationsBases}. These are familiar enough to be assumed without further comment in some practice-oriented actuarial syllabuses (see the quotation above). Moreover, each relationship has an illuminating corollary, as follows.

\begin{itemize}
\item[{\bf R1*}:] The contractual premium can be decomposed into a pure risk premium and two separate loadings for surplus: one capitalized at outset, the other emerging as premiums are paid.
\item[{\bf R2*}:] The EPV of the cumulative surplus at any time can be decomposed into: (a) initial surplus, plus; (b) premium loadings, plus; (c) a sum of pairs, each pair consisting of a systematic part and a martingale residual, for each source of surplus.
\item[{\bf R3*}:] The EPV of the total surplus is independent of the valuation basis (mentioned as {\bf R3 (iii)} above).
\end{itemize}

{{\bf R2} and {\bf R3*} are standard in a discrete-time setting with two technical bases, see for example \cite{olivieri2015}.} {The research-oriented literature addresses {{\bf R1}, {\bf R2} and {\bf R2*}} in the context of two technical bases, but not {{\bf R1*}, {\bf R3} or {\bf R3*}}, or the system as a whole}. This leads us to the {three}  purposes of this paper.

\begin{bajlist}

\item We set out to repair the gaps described above, in the setting of Markov models. That is, to describe {three technical bases}, the relationships {\bf R1} to {\bf R3} {and their corollaries {\bf R1*} to {\bf R3*}} in quite general terms.


\item We may say that a technical basis becomes operational by parametrizing Thiele's equation(s). We propose then to classify technical bases in terms of the boundary conditions assumed; initial, terminal or both. This leads naturally to the relationships {\bf R1} to {\bf R3} {and their corollaries}, and extends to Markov models. Along the way, we suggest retiring the classical retrospective policy value, which has never found an agreed definition in general Markov models (see the Appendix).

\item We also have in mind that different terminologies (first-order, second-order {\em versus} premium, valuation, experience), techniques (net premium {\em versus} gross premium),  frameworks (discrete-time {\em versus} continuous-time) and even pedagogies, can hamper communication between actuaries brought up in different traditions. We hope that this paper may help to bridge any gaps.

\end{bajlist}

{A study which shares some of these aims is \cite{moeller2007}, in particular Chapters 2 and 6. This also breaks away from Scandinavian-style regulation, but in a different way, aiming to base life insurance liabilities on retrospective considerations. They also consider the analysis of technical surplus into systematic and martingale components, {{\bf R2*} above}, begun by \cite{ramlau-hansen1988a} and since axiomatized by \cite{schilling2020} {and \cite{jetses2022}}. We will compare approaches in the extended example of Section \ref{sec:OtherExample}.}



\subsection{The Literature on Thiele's Equation, Markov Models  and Surplus}
\label{sec:IntroDevelopment}

The earlier history of Thiele's equation is told in \cite{hoem1983} and \cite{norberg2004}. Notably, its originator, the Danish actuary Thorvald Thiele, never published it, and it first appeared in print in \cite{gram1910} and \cite{joergensen1913} (though it is proved {\em en passant} in deriving Lidstone's theorem \cite[p.227]{lidstone1905}). 

Its second phase of development began with Hoem showing that it generalized easily to Markov models in continuous time, parametrized by transition intensities, representing a wide range of possible insurance contracts, see \cite{hoem1969, hoem1988}, {and also \cite{sverdrup1965} for a notable antecedent}. 

Following Hoem, many papers, in particular \cite{norberg1985, norberg1990, norberg1991, norberg1992, norberg1995} and others (\cite{ramlau-hansen1988a, wolthuis1986, wolthuis1987, wolthuis1990, wolthuis1992, wolthuis1994, milbrodt1993, milbrodt1997}) pursued multiple-state models. Emergence of surplus was a major theme (\cite{ramlau-hansen1988b, linnemann1993, linnemann1994, linnemann1995, linnemann2003}) followed by distribution of bonus (\cite{width1986, ramlau-hansen1991, norberg1999, linnemann2003, linnemann2004}). 

\cite{hoem1978} were first to translate life insurance mathematics into counting process language. It was by no means adopted uniformly in all the work cited above, but its power to bring new insights was perhaps shown particularly by \cite{ramlau-hansen1988a, ramlau-hansen1988b} on second moments and surplus, and \cite{norberg1991, norberg1992} on the r\^ole of filtrations as a model of information. {We already mentioned \cite{moeller2007} as a study that takes these ideas further, in models with more than two technical bases, see Section \ref{sec:OtherExample} below.}

Most recently, research has moved towards combined models of financial risk (canonical model, Black-Scholes) and long-term insurance risk (canonical model, life insurance), though without yet arriving at any clear destination. This lies beyond our scope. {For more details see}, for example, \cite{moeller2007}.


\subsection{Plan of this Paper}
\label{sec:Plan}

Section \ref{sec:echValnBases} introduces basic ideas in the context of the simplest life insurance contract; Thiele's differential equation, evaluating past and future cashflows, and technical bases. These ideas form the basis of Section \ref{sec:TechBase} where the Markov model is introduced, with particular attention to the counting process representation of the data, and the special r\^ole of the {`true' or `experience'} technical basis. 

Section \ref{sec:ClassificationOverall} proposes a classification of technical bases in terms of the boundary conditions satisfied in the associated Thiele's equation. To an extent this replaces the classical retrospective policy value as a model of accumulation depleted by decremental forces; the retrospective policy value is discussed further in the Appendix. The analysis is general enough to cover a wide range of valuation methods and extends the literature beyond  Scandinavian-style `first-order' and `second-order' technical bases.

Section \ref{sec:ModelsOutOfBases} introduces a model for the balance sheet of a life insurer in terms of three technical bases, and analyzes the surplus arising during a contract's term. The analysis is expressed as three relationships between pairs of technical bases, each well-known in practice, and extends the analysis of \cite{ramlau-hansen1988a}, {\cite{schilling2020} and \cite{jetses2022}}. Finally, Section \ref{sec:Examples} gives two {applications}; lapse-supported business {and {a model proposed by} \cite{moeller2007}}. 

Our conclusions are in Section \ref{sec:Concs}.


\section{A Simple Life Insurance Model}
\label{sec:echValnBases}

\subsection{Thiele's Differential Equation}
\label{sec:ThieleEquation}

Before introducing the Markov model, consider the simplest life insurance policy, commencing at age $x$, with term $n$ years. On death at time $t$ within the term a sum insured of $S$ is paid immediately, and on survival for $n$ years a maturity benefit of $M$ is paid (possibly $M=0$). Level premiums are paid {continuously} throughout the term at rate $P$ per year. We {assume there are no lump-sum cashflows, except benefits,} and ignore expenses to keep notation simple.

Suppose we are given a technical basis consisting of a force of interest $\delta_t$ and force of mortality $\mu_t$. We suppress the initial age $x$, so $\mu_t$ is understood to mean  $\mu_{x+t}$. Then Thiele's differential equation {\citep{dickson2020}} is:

\begin{equation}
\frac{d}{dt} \, V_t = \delta_t \, V_t + P - \mu_{t} \, (S - V_t). \label{eq:Thiele}
\end{equation}

\noindent The function $V_t$ of time $t$ represents a fund growing at force of interest $\delta_t$, while being added to by a continuous stream of premiums at rate $P$ per year, and decremented by a force of mortality $\mu_{t}$ that causes a sum insured of $S$ to be paid immediately on death, this cost to the insurer being offset by the reserve $V_t$ they hold.


\subsection{A Model of a Life Insurance Fund: One Technical Basis}
\label{sec:ModelFund}

Define the discount factor allowing for survivorship as well as interest:

\begin{equation}
\varphi_t = \exp \left( - \int_0^t ( \delta_s + \mu_{s}) \, ds \right) \label{eq:IntegratingFactor}
\end{equation}

\noindent then write down the equation of value solved to determine $P$:

\begin{equation}
0 = V_0 = \int_0^n \varphi_s \, (\mu_{s} \, S - P) \, ds + \varphi_n \, M. \label{eq:EqPrin}
\end{equation}

\noindent Split this integral at {any} time $t$, divide by $\varphi_t$ and rearrange:

\begin{equation}
\underbrace{\int_0^t \frac{\varphi_s}{\varphi_t} \, ( P - \mu_{s} \, S ) \, ds}_{\mbox{\scriptsize Retrospective Policy Value}} = \underbrace{\int_t^n \frac{\varphi_s}{\varphi_t} \, ( \mu_{s} \, S - P ) \, ds + \frac{\varphi_n}{\varphi_t} \, M}_{\mbox{\scriptsize Prospective Policy Value}}. \label{eq:EqPrinPrem2}
\end{equation}

\noindent We can verify that the left-hand side of equation (\ref{eq:EqPrinPrem2}) is the solution of Thiele's equation with initial boundary condition $V_0=0$, and the right-hand side is the solution with terminal boundary condition $V_n=M$.

We therefore have, in a single equation (\ref{eq:EqPrin}), a mathematical model of the evolution of assets and liabilities under a life insurance policy, such that:

\begin{bajlist}
\item {there is a single technical basis for premiums, policy values and the experience};
\item the premium rate $P$ satisfies the equivalence principle;
\item no capital is required at outset ($V_0=0$);
\item no profit remains after maturity ($V_n=M$); 
\item all policy values are solutions of Thiele's equation; and
\item retrospective and prospective policy values are equal {at all times $t$} (a kind of self-financing condition).
\end{bajlist}

\noindent \cite{norberg1991} noted: ``Thus, the traditional concept of `retrospective reserve' is rather a retrospective formula for the prospective reserve $\ldots$''.


\subsection{A Stochastic Decomposition}
\label{sec:StochDecomp}

We get more insight into equation (\ref{eq:EqPrinPrem2}) following \cite{norberg1991} by writing down the cumulative cashflow as a stochastic process and considering the information acquired by observing {this} process up to time $t$ (see the Appendix (e) for more details). Let $A(t)$ be the total cashflow up to time $t$ (for example, in Section \ref{sec:ModelFund}, $A(t) = \int_0^t (\mu_s S - P) \, ds$), and let $v(t)$ be a discount function (for example, $v(t) = v^t$). Then the value at time $t$ of all cashflows, denoted by $V(t)$, again splitting the integral at time $t$, is:

\begin{equation}
V(t) = - \frac{1}{v(t)} \int_{[0,t]} v(r) \, d(-A)(r) + \frac{1}{v(t)} \int_{(t,n]} v(r) \, dA(r). \label{eq:Split1}
\end{equation}

\noindent Taking expectations conditional on ${\cal F}_t$, information available at time $t$, we get:

\begin{equation}
\Mean[V(t) \mid {\cal F}_t] = - \frac{1}{v(t)} \int_{[0,t]} v(r) \, d(-A)(r) + \frac{1}{v(t)} \, \Mean \left[ \int_{(t,n]} v(r) \, dA(r) \, \Big| \, {\cal F}_t \right]. \label{eq:Split}
\end{equation}

\noindent We recognize the first term on the right-hand side as the accumulation of actual cashflows, and the second term as the prospective policy value. Taking expectations of equation (\ref{eq:Split}) conditional on ${\cal F}_0$, we get:

\begin{equation}
\Mean[V(t) \mid {\cal F}_0] = \underbrace{- \frac{1}{v(t)} \Mean \left[ \int_{[0,t]} v(r) \, d(-A)(r) \, \Big| \, {\cal F}_0 \right]}_{\mbox{\scriptsize Retrospective Policy Value}} + \underbrace{\frac{1}{v(t)} \, \Mean \left[ \int_{(t,n]} v(r) \, dA(r) \, \Big| \, {\cal F}_0 \right]}_{\mbox{\scriptsize Prospective Policy Value}}. \label{eq:Split2}
\end{equation}

\noindent If the equivalence principle holds, this is zero, {and the insurance contract is, in expectation, a self-financing portfolio, in the derivative-pricing sense (see, for example, \cite{baxter1996})}. We see that the conventional meaning of `accumulation' in life insurance mathematics is in fact an expected value conditional on ${\cal F}_0$, {a fact to add to Norberg's comment in Section \ref{sec:ModelFund} above.} 


\subsection{Technical Bases and Regulations}
\label{sec:Technical Bases}

Motivated by the above, we define a {\em technical basis}, denoted by ${\cal B}$, to be the pair ${\cal B} = (\delta_t,\mu_t)$. We then ask if one technical basis can serve all purposes, at all times, and answer `no', as the following examples show.

\begin{bajlist}

\item As time passes, the interest and mortality used to calculate the contractual premium fade into historic more than current interest. 

\item The valuation of past cashflows may be a matter of factual accounting; the valuation of future cashflows rarely is.


\end{bajlist}

Under Scandinavian-style regulation, the first-order  technical basis serves for both premiums and valuations. There is one technical basis ${\cal B}$, one premium rate, denoted by $P$, which satisfies equation (\ref{eq:EqPrin}) under ${\cal B}$ and $P$ is also the premium valued, meaning that when we write a policy value as:

\begin{equation}
\mbox{Policy Value} = \mbox{EPV[Future Benefits]} - \mbox{EPV[Future Premiums]} \label{eq:PolValFormula}
\end{equation}

\noindent the future premiums referred to are at rate $P$ per annum. 

Under other valuation regimes this restriction on the choice of technical basis {\em and} the choice of premium valued may be lifted (see \cite{fisher1965} for example).

\begin{bajlist}

\item There may be separate technical bases for premiums and valuations, which we may denote by ${\cal B}^P$ and ${\cal B}^L$ respectively. This may be mandatory or merely permissive. 

\item Let $P$ be the premium rate under ${\cal B}^P$ --- that is, the premium rate $P$ that solves equation (\ref{eq:EqPrin}) under ${\cal B}^P$ --- and let $\pi^*$ be the corresponding premium rate under ${\cal B}^L$. Then we have a choice as to the premium rate actually valued --- that is, the premium rate defining the cashflow in the last term in equation (\ref{eq:PolValFormula}).

\begin{bajsublist}

\item If $\pi^*$ is the premium rate valued we have a {\em net premium valuation}, and it will still be true that $V_0=0$.

\item If $P$ is the premium rate valued we have a {\em gross premium valuation}, and in general it will no longer be the case that $V_0=0$.

\item The rules may be still broader, however, and allow the actuary to choose a valuation premium rate different from both $P$ and $\pi^*$; denote this by $\pi^L$. This means that in a policy value of the form (\ref{eq:PolValFormula}) we calculate EPVs using ${\cal B}^L$ but the premium rate valued is $\pi^L$. 

\end{bajsublist}

\noindent Hence we need both ${\cal B}^L$ and $\pi^L$ to specify policy values, see Section \ref{sec:DefsTBandPP}.

\item In practice, the choice of valuation basis may be even wider than shown above. For example, \cite[p.269]{fisher1965} suggests in some {special} cases using a policy value equal to a few years' net premiums, perhaps accumulated at interest. In such cases, policy values need not be solutions of Thiele's equation. In this paper, we restrict attention to policy values that are solutions of Thiele's equation.

\end{bajlist}

{We summarize below how the elements of a regulatory regime may differ from the Scandinavian style.

\begin{bajlist}

\item Under Scandinavian-style regulations, ${\cal B}^L = {\cal B}^P$ and $\pi^L = \pi^* = P$. Policy values are always solutions of Thiele's equation. Everything simply follows the first-order technical basis. These {\em constraints} are so familiar to those accustomed to them that they often go unstated.

\item Under other regulatory regimes, we may have ${\cal B}^L \not= {\cal B}^P$ and $\pi^L$ equal to $P$ (gross premium valuation) or $\pi^*$ (net premium valuation), or neither of $P$ or $\pi^*$ (actuary's discretion). Policy values are usually solutions of Thiele's equation but need not be. These {\em freedoms} are so familiar to those accustomed to them that they often go unstated.

\end{bajlist} 

\noindent If this paper has a communications element, it consists of stating clearly what is described above as often going unstated.} 

In the next section, we formalize the intuitive discussion above.

\subsection{Valuation Bases, Pure Premiums, and Valuation Premiums}
\label{sec:DefsTBandPP}

Given an arbitrary technical basis ${\cal B}$, define the associated {\em pure premium rate}, denoted by $\pi^*({\cal B})$, to be the premium rate that satisfies the equation of value under ${\cal B}$. That is, $\pi^*({\cal B})$ {is chosen} such that:

\begin{equation}
0 = V_0 = \int_0^n \varphi_s \, (\mu_{s} \, S - \pi^*({\cal B})) \, ds + \varphi_n \, M \label{eq:EqPrinPurePrem}
\end{equation}

\noindent {where $\mu_t$ and $\varphi_t$ are on technical basis ${\cal B}$}. Note that we are assuming, without further comment, that the equation of value under ${\cal B}$ has a unique `sensible' solution\footnote{In general the equation of value has many solutions (it is often a polynomial equation, for example). What we require is that it has a unique real and non-negative solution.} $\pi^*({\cal B})$.

{Let ${\cal B}^P, {\cal B}^L$ and ${\cal B}^A$ be the technical/valuation bases intended for the calculation of premiums, policy values and accumulations respectively. Define the {\em contractual premium} $P$ to be $P =  \pi^*({\cal B}^P)$ {and the {\em valuation pure premium} $\pi^*$ to be $\pi^* = \pi^*(\tilde{\cal B}^L)$}.

{The fact that ${\cal B}^L$ is the valuation basis means that it is the basis used to calculate the EPV of future cashflows. It does not mean that the valuation pure premium $\pi^*$ is necessarily one of those future cashflows, namely the premium rate valued, see Section \ref{sec:Technical Bases}.} Therefore we introduce an {\em augmented technical basis} $\tilde{\cal B}^L = ({\cal B}^L , \pi^L)$ consisting of the technical basis ${\cal B}^L$ and the premium rate actually to be valued $\pi^L$. We define $\pi^*(\tilde{\cal B}^L)=\pi^*({\cal B}^L)$, and we define a new function $\pi(\tilde{\cal B}^L)=\pi^L$. The only requirement we impose is that $\pi^L$ is the pure premium rate associated with {\em some} technical basis ${\cal B}'$, that is $\pi^L = \pi^*({\cal B}')$, {so that the resulting policy values are solutions of Thiele's equation}. By choosing $\pi^L$ appropriately, we can specify net premium, gross premium or other valuation methods.  

\begin{bajlist}
\item Choose $\pi^L = \pi^*({\cal B}^P) = P$ and we have a gross premium valuation basis.
\item Choose $\pi^L = \pi^*({\cal B}^L) = \pi^*$ and we have a net premium valuation basis.
\item Choose ${\cal B}^L = {\cal B}^P$ and $\pi^L = P$; then the valuation basis is the same as the premium basis and we have Scandinavian-style regulation.
\end{bajlist}

\noindent These cases probably cover the main examples of interest. However, $\pi^L$ is arbitrary within sensible limits, and is not confined to the examples above. {A case in point might be a net premium valuation  basis different from the premium basis, but with a maximum premium valued equal to 90\% of $P$. Then for some contracts $\pi^L \not= \pi^*({\cal B}^L)$ and $\pi^L \not= P$.}

We call ${\cal B}^A$ the {\em accumulation basis}, and for simplicity we assume all accumulations of past cashflows  accumulate the contractual premium $P$.

{Table \ref{table:PremiumRates} summarizes the main premium rates defined in respect of technical bases.}

\begin{table}
\begin{center}
\caption{\label{table:PremiumRates} Main premium rates defined in respect of technical bases.}
\small
\begin{tabular}{cl}
& \\
Rate & Description \\[0.5em]
$\pi^* = \pi^*({\cal B})$ & Pure premium rate on technical basis ${\cal B}$ \\
$\pi^L = \pi(\tilde{\cal B}^L)$ & Valuation premium rate {specified by augmented valuation basis $\tilde{\cal B}^L$} \\
$P = \pi^*({\cal B}^P)$   & Contractual premium rate on premium basis ${\cal B}^P$.
\end{tabular}
\end{center}
\end{table}


\section{Markov Models, Data and Thiele's Equations}
\label{sec:TechBase}




\subsection{Definitions: Markov Models}
\label{sec:DefsModels}

Suppose a process $J(t), (0 \le t \le n)$ takes values in a state space ${\cal S} = \{1, 2, \ldots, m\}$, labelling states representing `alive', `ill', `dead' and so on. Define $J(t)$ to be the state occupied by the life at time $t \ge 0$, age $x+t$; we assume $J(t)$ is right-continuous with left-hand limits, and $J(0)=1$. Define {\em occupancy probabilities}, denoted by $P_{ij}(t,s)$, as:

\begin{equation}
P_{ij}(t,s) = \Prob[ \, J(t+s) = j \mid J(t) = i \,] \qquad (i, j \in {\cal S}) \label{eq:OccProbs}
\end{equation}

\noindent and {\em transition intensities}, denoted by $\mu_t^{ij}$, as:

\begin{equation}
\mu_t^{ij} = \lim_{h \to 0^+} \frac{1}{h} \, \Prob[ \, J(t+h) = j \mid J(t) = i \, ] \qquad (i, j \in {\cal S}, i \not= j) \label{eq:Intensity}
\end{equation}

\noindent assuming all such limits exist. The Markov assumption is present in conditioning only on $J(t)=i$, the state occupied at time $t$, excluding any other history. Also define $\mu_t^i = \sum_{j \not= i} \mu_t^{ij}$ and the probabilities of not leaving a state $i$ are defined as:



\begin{equation}
\bar{P}_i(t,s) = \Prob[ \, J(t+r) = i, 0 \le r \le s \mid J(t) = i \,] = \exp \left( -\int_t^{t+s} \mu^i_r \, dr \right). \label{eq:SurvProbs}
\end{equation}


\subsection{Definitions: Insurance Contracts}
\label{sec:DefsContracts}

Under a multiple-state model, insurance contracts are defined by payments of three types, namely lump sums paid immediately on a transition between states, lump sums payable upon expiry in a given state, and premiums paid continuously during a sojourn in a state. Accordingly define:

\begin{eqnarray*}
b_{ij} & = & \mbox{Lump sum paid on transition from state $i$ to state $j$} \\
M_i & = & \mbox{Lump sum paid on expiry while in state $i$} \\
P_i & = & \mbox{Rate of premium payable continuously during sojourn in state $i$}.
\end{eqnarray*}

\noindent Cashflows from the insurer to the insured are positive, premiums are treated as a negative annuity, and so on. For simplicity we have made the following assumptions, all of which can be relaxed at the cost only of more notation.

\begin{bajlist}

\item We ignore annuity-type benefits.

\item We consider only constant benefits, but they can be made functions of time. 

\item We ignore expenses.

\end{bajlist}

If it helps, we can represent a contract by the collection of its parameters, denoted by ${\cal C}$, thus: ${\cal C} = (b_{ij},M_i,P_i)$. We suppress other factors such as age, term, sex and so on, that can be included as required by the  context. 


\subsection{Representations of Multiple-State Model Data}
\label{sec:DefsRepresentations}

\begin{bajlist}

\item A sample path of $J(t)$ is a piecewise-constant function on $[0,n]$, taking values in $\{1, 2, \ldots, m\}$, right-continuous with left-hand limits. We call this the {\em sample-path} representation of the process. 

\item Another representation generalizes the idea of the time until death of a life aged $x$ being the random variable $T_x$  (\cite{bowers1997, gerber1990, olivieri2015, dickson2020}). A `sample point' is now a random integer $k \ge 0$, a  sequence of random times $0 < T_1 < T_2 \ldots, T_k \le n$ and, for each random time $T_j$, a pair of states called a {\em mark}, identifying a transition in the model. This is the {\em marked point process} (MPP) representation of the process. 

\item A third representation is in terms of {\em counting processes}. Define $N^{ij}_t$ to be the number of direct transitions from state $i$ to state $j$ in $[0,t]$. Then $N_t$ defined as the collection $\{ N^{ij}_t : i,j \in {\cal S}, i \not= j \}$, with the condition that no two processes in the collection can jump simultaneously, is a multivariate counting process. 

\end{bajlist}

\noindent These representations are completely equivalent, see \cite[Chapter 2]{jacobsen2006} for technical details. Also, they describe the {\em data generated by the process we are modelling}, not the models themselves. To see this, note that none of them mentions the intensities $\mu_t^{ij}$.

Useful with all three representations are the following  {\em indicator processes} of presence in states, denoted by $Y^i_t$, for $i \in {\cal S}$:

\begin{equation}
Y^i_t = I_{\mbox{\scriptsize \{$J(t^-)=i$\}}}= I_{\mbox{\scriptsize \{In state $i$ at time $t^-$\}}}, \label{eq:YDef}
\end{equation}

\noindent where $I_{A}$ is the usual indicator of event $A$. $Y^i_t=1$ means that the process $N_t$ is at risk of jumping out of state $i$ at time $t$. 

Each representation can also be equipped with a filtration (non-decreasing sequence of $\sigma$-algebras) constituting a model of {\em information} obtained by observing events over time. By an abuse of notation, let $\{{\cal F}_t\}_{t \ge 0}$ denote the filtration under any of the three representations. 

For more on counting processes, and the revolution they have wrought on statistical inference in survival models, see the standard texts \cite{andersen1993} or \cite{fleming1991}. 


\subsection{Technical Bases, Pure Premiums, and {Valuation Premiums}}
\label{sec:TechBaseDefs}

For simplicity, we assume a force of interest $\delta$ that is constant and the same for all states, hence we can write all discount factors as:

\begin{equation}
\exp \left( -\int_0^t \delta^i_r \, dr \right) = \exp \left( -\int_0^t \delta_r \, dr \right) = v^t 
\end{equation}

\noindent for $v = \exp(-\delta)$. The definitions then follow very much along the lines of Section \ref{sec:DefsTBandPP}.

\begin{bajlist}

\item A {\em technical basis} for a given Markov model and contract ${\cal C}$, is a collection ${\cal B} = (\delta,\mu_t^{ij})$ of force of interest and transition intensities. 

\item The {\em pure premium} rates associated with the technical basis ${\cal B} = (\delta,\mu_t^{ij})$ are the premium rates $\pi_i^*$ satisfying the equation of value:

\begin{equation}
0 = \sum_i \int_0^n v^s \, P_{1i}(0,s) \, \left[ \pi_i^* - \sum_{j \not= i} \mu_s^{ij} \, b_{ij} \right] \, ds - v^n \sum_i P_{1i}(0,n) \, M_i \label{eq:PurePrem}
\end{equation}

\noindent and other constraints (see below) that ensure uniqueness, and it is denoted by $\pi^* = \pi^*({\cal B})$, with $i$th element $\pi^*_i = \pi^*({\cal B})_i$. 

\item {Exactly as in Section \ref{sec:DefsTBandPP}, for valuation the technical basis ${\cal B}^L$ is augmented with a set of premium rates $\pi_i^L$ called the {\em valuation premium rates} to define the {\em valuation basis} $\tilde{\cal B}^L$, and we define the functions $\pi^*(\tilde{\cal B}^L)$ and $\pi(\tilde{\cal B}^L)$ analagously. We leave the details to the reader.} 

\item Given a technical basis, we require some constraints on the admissible values of the premium rates $\pi^*_i$ to ensure that the equation of value (\ref{eq:PurePrem}) has a {unique solution. A simple example would be that level premiums are payable continuously while in state 1, and no premiums are payable otherwise. Other sensible solutions may exist, such as premiums increasing linearly over time, so we assume constraints force a choice among all such solutions of equation (\ref{eq:PurePrem}), so the function $\pi^*({\cal B})$, defined above, makes sense.}

\end{bajlist}


\subsection{Prospective Policy Values and Thiele's Equations in a Markov Model}
\label{sec:ThieleMarkov}

Assume a force of interest $\delta$, and premium rates $\pi_i$ payable while in state $i$. \cite{hoem1969, hoem1988} defined prospective policy values in state $i$, denoted by $V^i_t$, as (our notation):

\begin{equation}
V_t^i = \int_0^{n-t} v^s \, \sum_{j=1}^m P_{ij}(t,s) \, \bigg[ - \pi_j + \sum_{k \not= j} \mu_{t+s}^{jk} \, b_{jk} \bigg] \, ds + v^{n-t} \, \sum_{j=1}^m P_{ij}(t,n-t) \, M_j. \label{eq:ProspPolValMarkov}
\end{equation}

\noindent Thiele's equations for policy values are the system:

\begin{eqnarray}
\frac{d}{dt} \, V^i_t & = & \delta \, V^i_t + \pi_i - \sum_{j \not= i} \mu_t^{ij} \, (b_{ij} + V^j_{t} - V^i_{t}) \qquad (i \in {\cal S}) \label{eq:ThieleMarkov1} \\
& = & \delta \, V^i_t + \pi_i - \sum_{j \not= i} \mu_t^{ij} \, R_t^{ij} \qquad (i \in {\cal S}) \label{eq:ThieleMarkov}
\end{eqnarray}

\noindent where $R_t^{ij} = b_{ij} + V^j_{t} - V^i_{t}$, allowing for a new policy value to be set up on a transition between states. 

\newpage

\section{A Classification of Technical Bases}
\label{sec:ClassificationOverall}


\subsection{Motivation}
\label{sec:ClassificationMotivation}

Conventionally, technical bases have parametrized Thiele's equations, which in turn defined policy values as solutions. Policy values were prospective or retrospective and the two were equal under `nice' circumstances, see Section \ref{sec:ModelFund}, which provided life insurance mathematics with a form of self-financing condition.

The awkward quantity in this scheme is the retrospective policy value. Prospective policy values have a clear purpose, in which the pooling of {\em future} stochastic risk is clearly {probabilistic and} defined in terms of transition intensities. Accumulations of past cashflows are also conceptually clear, {and not probabilistic}, but retrospective pooling of risk {is expressed in probabilistic language}, in terms of transition intensities --- see yet again Norberg's comment cited at the end of Section \ref{sec:ModelFund}, and Section \ref{sec:StochDecomp}, --- and, when we consider Markov models, apparently not well-defined --- see the Appendix.

We sidestep this difficulty in the following sections by defining an {\em accumulation fund} to be a solution of Thiele's equations satisfying an initial boundary condition of having value zero in every state; therefore, an original condition of no assets. The traditional retrospective policy value as used in the literature (in particular, in \cite{ramlau-hansen1988a}) will often coincide with this definition, but we do not rely on that being so.


\subsection{Proposed Classification of Technical Bases}
\label{sec:ClassificationProposed}

Our classification of technical bases, in terms of boundary conditions satisfied by solutions $V_t^i$ of the corresponding Thiele's equations, is set out below, and summarized in Table \ref{table:Classification}. {The general idea is that two out of the premium rate, the initial boundary condition and the terminal boundary condition are fixed, and the third is to be determined.} Note that every {policyholder} is assumed to start in state 1.

\begin{bajlist}

\item If the boundary conditions $V_0^i=0 \: (i \in {\cal S})$ are satisfied, we call ${\cal B}$ an {\em accumulation basis}, and the associated functions {$V_t^i$} {\em accumulation funds}, alternatively {\em policy accounts}. It represents the accumulation of a fund with no initial endowment. We often use the notation ${\cal B}^A$ and may denote the fund by $A_t^i$.

\item If the boundary conditions $V_n^i=M_i \: \: (i \in {\cal S})$, are satisfied, we call the {augmented technical basis $\tilde{\cal B} = ({\cal B},\pi^L)$ a {\em valuation basis}}, and the associated functions $V_t^i$ {\em policy values} or sometimes {\em prospective policy values}. We often use the notation {${\cal B}^L$ and $\tilde{\cal B}^L$. The premiums valued are $\pi(\tilde{\cal B}) = \pi^L$, see Section \ref{sec:TechBaseDefs}}.

\item {If both sets of boundary conditions $V_0^1=0$ and $V_n^i=M_i \: \: (i \in {\cal S})$ are satisfied}, we call $\tilde{\cal B} = ({\cal B},\pi^*({\cal B}))$ a {\em proper valuation basis}, and the associated functions $V_t^i$ are again called policy values. 

\item {Any proper valuation basis $\tilde{\cal B}$ is a candidate to be chosen as the {\em premium basis}; precisely one {\em must} be so chosen, and in that r\^ole it is denoted by ${\cal B}^P$.} 

\item If none of the above boundary conditions are met the technical basis defines a fund with no special name.

\end{bajlist}

\begin{table}
\begin{center}
\caption{\label{table:Classification} Names of technical bases, and associated functions $V_t$ or $A_t$ which are solutions of Thiele's equation, depending on boundary conditions satisfied. General idea is that two out of the premium rate, the initial boundary condition and the terminal boundary condition are fixed, and the third is to be determined. {Valuation basis $\tilde{\cal B}^L$ is an augmented technical basis, see Sections \ref{sec:DefsTBandPP} and \ref{sec:TechBaseDefs}}. {The premium basis ${\cal B}^P$ is necessarily also a proper valuation basis.} See Section \ref{sec:MartingaleBasis} for the asset share ${}^M \! \! A_t^i$, a special case of an accumulation fund.}
\small
\begin{tabular}{llll}
& & & \\
Basis & Boundary Condition(s) & Name of Basis & Name, Notation  of Function \\[0.5ex]
${\cal B}$   & None specified & Technical Basis & Fund, $V_t^i$ \\
${\cal B}^A$ & $A_0^i=0$ & Accumulation Basis & Accumulation Fund, or \\
& & & Policy Account, $A_t^i$ \\
$\tilde{\cal B}^L$ & $V_n^i=M_i$ & Valuation Basis & Policy Value, $V_t^i$ \\
${\cal B}^P$ & $V_0^1=0$ AND $V_n^i=M_i$ & {Premium Basis} & Policy Value, $V_t^i$ \\[0.5ex]
${\cal B}^M$ & ${}^M \! \! A_0^i=0$ & Experience Basis & Asset Share, ${}^M \! \! A^i_t$ 
\end{tabular}
\end{center}
\end{table}

{The different treatment of funds at time $t=0$ is because of their interpretation in the balance sheet. A non-zero policy value at outset signals a movement on the liability side, either the release of surplus or a need for capital support. It is important that these are realistic and transparent because they can be created or conjured away by choosing the valuation basis. A non-zero accumulation fund, on the other hand, would (or should) represent a movement on the assets side of the balance sheet and should not be an artifact of the technical basis.} {This does {\em not} say that assets must be taken at market value in the balance sheet.}


\subsection{Canonical Examples of Valuation Bases}
\label{sec:CommentsClassification}

The examples of valuation bases for the simple life insurance model (see Section \ref{sec:DefsTBandPP}) can be restated in terms of the Markov model.

\begin{bajlist}

\item Valuation bases $\tilde{\cal B}^L$ with $\pi(\tilde{\cal B}^L)=P$ are called {\em gross premium valuation bases}.

\item Valuation bases $\tilde{\cal B}^L$ with $\pi(\tilde{\cal B}^L) = \pi^*({\cal B}^L)$ are called {\em net premium valuation bases} (and must also be proper valuation bases).

\item {Under Scandinavian-style regulations, the first-order technical basis defines both a gross premium and a net premium valuation basis. Most of the literature uses such a basis.}

\end{bajlist}


\begin{figure}
\begin{center}
\includegraphics[scale=0.87]{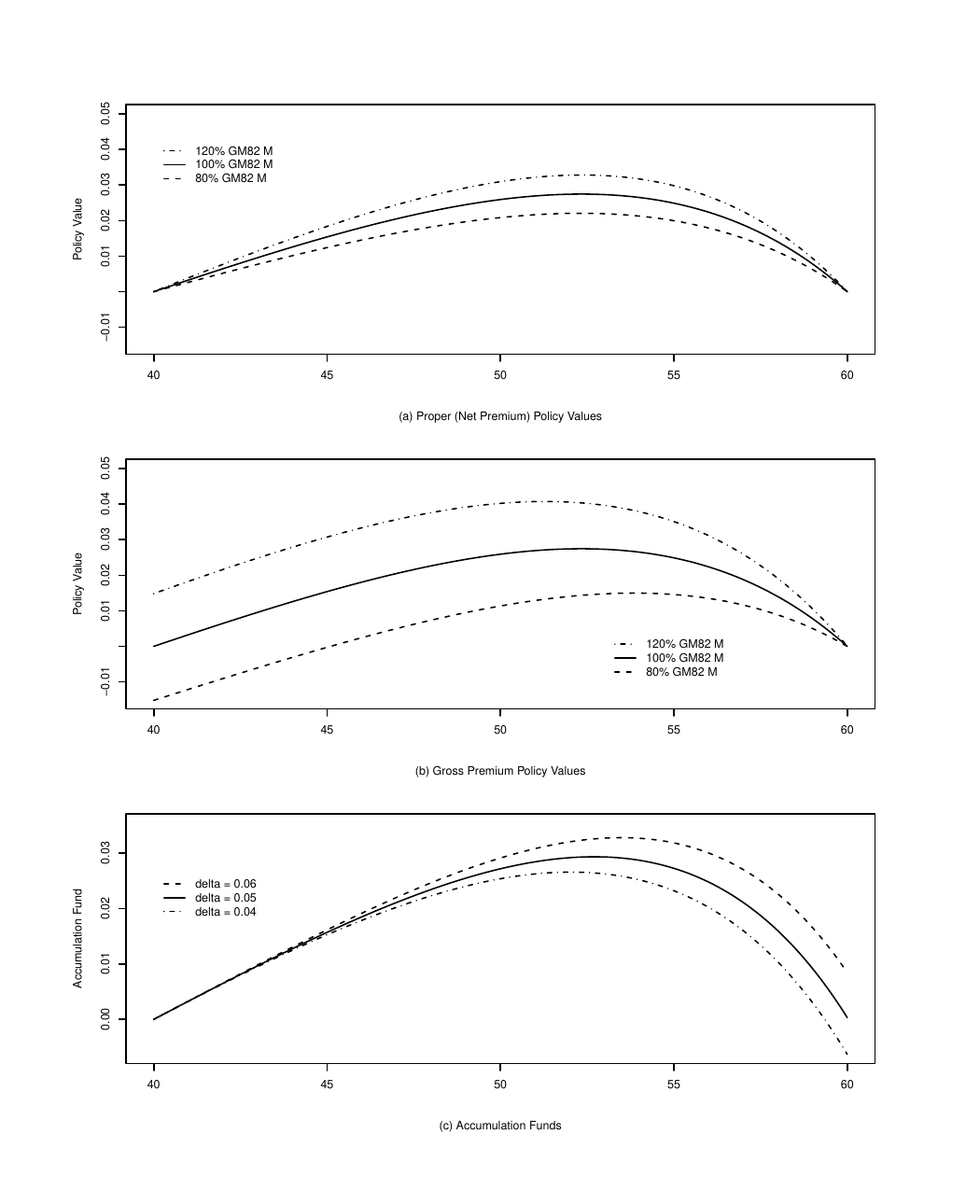}
\caption{\label{fig:TechBases} Examples of policy values and accumulation funds. Term insurance, sum assured \$1, age 40, term 20 years. Baseline force of interest $\delta=0.05$ and mortality $\mu_t$ = GM82 Males (Denmark). Premium basis ${\cal B}^P = (0.05, \mu_t, P=0.0063067)$. Panel (a) shows proper (net premium) policy values with mortality 120\%, 100\% and 80\% of baseline: e.g. $\tilde{\cal B}^L = (0.05, 0.8 \mu_t, \pi^*(\tilde{\cal B}^L))$. Panel (b) shows gross premium policy values with mortality 120\%, 100\% and 80\% of baseline: e.g. $\tilde{\cal B}^L = (0.05, 0.8 \mu_t, P)$. Panel (c) shows accumulation funds with $\delta = 0.04, 0.05$ and 0.06, e.g. ${\cal B}^A = (0.04,\mu_t)$.}
\end{center}
\end{figure}


\subsection{Examples}
\label{sec:PolValsExamples}

Figure \ref{fig:TechBases} illustrates policy values and accumulation bases for a term insurance contract (age 40, term 20 years, sum insured \$1), given a premium basis ${\cal B}^P = (0.05,\mu_t,P=0.0063067)$ where $\mu_t$ is on the Danish GM82 Males life table. Panel (a) shows proper policy values (equivalently, net premium policy values, see Section \ref{sec:CommentsClassification}), with both $V_0$ and $V_n$ fixed, but pure (valuation) premium $\pi^*(\tilde{\cal B}^L)$ varying; panel (b) shows gross premium policy values with $V_n$ and valuation premium $P$ fixed but $V_0$ varying; and panel (c) shows accumulation funds with $V_0$ and contractual premium $P$ fixed and $V_n$ varying. {Essentially, Scandinavian-style regulation of liabilities admits only a fixed technical basis represented by (a), net premium valuations a choice of technical bases represented by (a), and gross premium valuations also technical bases represented by (b).}


\subsection{The Experience Basis, Asset Shares and Martingales}
\label{sec:MartingaleBasis}

{We assume the existence of a unique accumulation basis ${\cal B}^{M} = (\delta,{}^M \! \mu_t^{ij})$ representing the actual experience and called the {\em experience} basis. It may also sometimes be called the {\em real}, {\em asset share} or {\em martingale} basis}. The corresponding fund, denoted by ${}^M \! \! A_t^i$, is called the {\em asset share}. 

A precise interpretation of the statement that technical basis ${\cal B}^{M}$ is `true', insofar as it concerns the intensities, is that the processes $M_t^{ij}$ defined by:

\begin{equation}
M^{ij}_t = N_t^{ij} - \int_0^t Y_s^{i} \, {}^{M} \! \mu_s^{ij} \, ds \label{eq:Martingale}
\end{equation}

\noindent {(see Section \ref{sec:DefsRepresentations} for definitions of $N_t^{ij}$ and $y_t^i$)} are a set of orthogonal ${\cal F}_t$-martingales {under ${\cal B}^M$}, and ${\cal B}^M$ is unique in this respect (this is the Doob-Meyer Decomposition {under ${\cal B}^M$}, see \cite{andersen1993} or \cite{fleming1991}). 

{Technical basis ${\cal B}^{M}$ occupies a privileged position in the model. It is the technical basis that is assumed to generate the data. With the counting process representation, it  provides a fundamental link between model and data. A technical basis with intensities estimated from the data strictly ought to be denoted by ${\cal B}^{\hat{M}}$ but we do not use this notation in the sequel.}


\subsection{A Note on Expected Values}
\label{sec:ExpectedValues}

It was assumed in Section \ref{sec:TechBase} that a technical basis defines transition probabilities $P_{ij}(t,s)$ (see equation (\ref{eq:OccProbs})), hence also an expected value operator, at least with respect to the transition intensities of the technical basis. Given technical basis ${\cal B}$ in this section, we could define an expected value operator $\Mean_{\cal B}[X]$. For the avoidance of doubt, we do not require this degree of abstraction. All expected values in this paper are based on the technical basis ${\cal B}^M$, and can (loosely) be regarded as `true' expectations {under the `real world' measure}, and we use the unadorned operator $\Mean[X]$ throughout to mean $\Mean_{{\cal B}^M}[X]$.


\subsection{A Note on Policy Accounts}
\label{sec:AccBasesUses}

An accumulation basis may be used to represent amounts credited to a policy in a {\em policy account}, which may subsequently be used as a basis for declaring bonus. See for example \cite{ramlau-hansen1988a}, {\cite{linnemann2003},} \cite{moeller2007}. Some smoothing of investment returns may be desired, and pooling of mortality risk is required, so the `raw' experience is not suitable for this purpose. Other parties may be due a share of any surplus, for example shareholders or the insurer's `estate', so the full {experience} basis ${\cal B}^M$ may not be appropriate either. Therefore, in what follows, we usually consider surplus with respect to an arbitrary accumulation basis ${\cal B}^A$.


\section{Surplus}
\label{sec:ModelsOutOfBases}

\subsection{A Simple Model of a Life Insurer's Balance Sheet}
\label{sec:ModelBalanceSheet}

The simplest model of a life insurer's balance sheet has one technical basis, see Section \ref{sec:ModelFund}, and attains a certain level of mathematical perfection, but does not support essential features from practice, such as surplus.  Most of the literature uses a model with `first-order' and `second-order' technical bases based on Scandinavian-style regulation.  We consider a model with three technical bases, one  of each type defined in Section \ref{sec:ClassificationOverall}. 

Our model is canonical, in the sense that: (a) models with fewer technical bases lose functionality; and (b) adding more technical bases duplicates functionality.

{As before}, we call the three bases the {\em premium basis}, {\em valuation (or policy value) basis} and {\em accumulation basis}\footnote{In the UK, historically, life insurance regulation devolved great responsibility for solvency reporting on the individual actuary. Technical bases were not imposed, and premium rates were not regulated. See \cite{Cox1962} or \cite{turnbull2017} for example, for accounts of these developments.}, denoted by ${\cal B}^P, \tilde{\cal B}^L$ and ${\cal B}^A$ respectively. If ${\cal B}^A = {\cal B}^{M}$, we have a model of asset shares. Otherwise, we have a model with an arbitrary policy account as asset. 

{Figure \ref{fig:RelationsBases2} illustrates the relationships between the three technical bases. It is the same as Figure \ref{fig:RelationsBases} except that `Experience Basis' has been replaced by `Accumulation Basis' since ${\cal B}^A$ is not necessariy the same as ${\cal B}^M$. For example, the accumulated quantity may be a policy account (Section \ref{sec:AccBasesUses}) which is credited with the fund rate of return less one percent.}


\begin{figure}
\begin{center}
\begin{picture}(100,65)
\put(50,60.5){\makebox(0,0)[c]{PREMIUM}}
\put(50,55.5){\makebox(0,0)[c]{BASIS}}
\put(15,10.5){\makebox(0,0)[c]{VALUATION}}
\put(15,5.5){\makebox(0,0)[c]{BASIS}}
\put(85,10.5){\makebox(0,0)[c]{ACCUMULATION}}
\put(85,5.5){\makebox(0,0)[c]{BASIS}}
\put(32.5,33){\vector(1,1){15}}
\put(32.5,33){\vector(-1,-1){15}}
\put(67.5,33){\vector(1,-1){15}}
\put(67.5,33){\vector(-1,1){15}}
\put(50,8){\vector(1,0){14}}
\put(50,8){\vector(-1,0){14}}
\put(20.0,35){\makebox(0,0)[c]{{\bf R1}: Surplus}}
\put(20.0,31){\makebox(0,0)[c]{$t = 0$}}
\put(79.5,35){\makebox(0,0)[c]{{\bf R3}: Profit}}
\put(80.5,31){\makebox(0,0)[c]{$t = n$}}
\put(50,4){\makebox(0,0)[c]{{\bf R2}: Surplus Rate}}
\put(50,0){\makebox(0,0)[c]{$0 \le t \le n$}}
\end{picture}
\end{center}
\caption{\label{fig:RelationsBases2} Relationships between the premium basis, valuation (policy value) basis and accumulation basis.}
\end{figure}
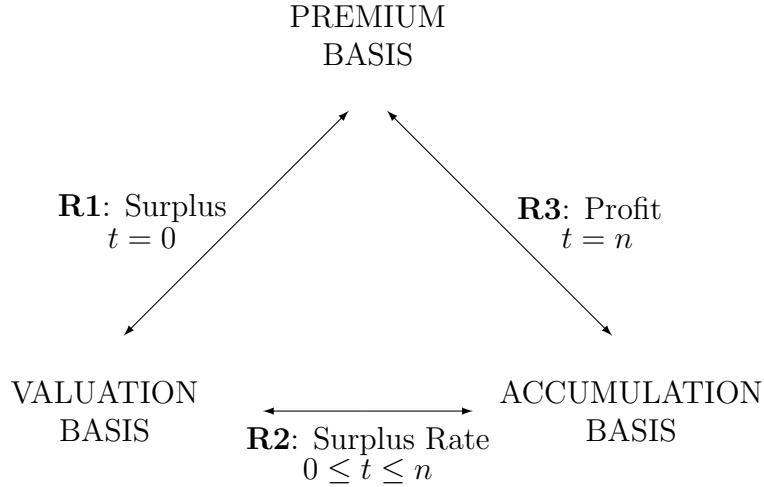


\subsection{Basic Surplus Relationship}
\label{sec:SurplusRelationships}

{Conventionally,} given a valuation basis $\tilde{\cal B}^L = ((\delta^L,{}^L\mu_t^{ij}),\pi_i^L)$ and an accumulation basis ${\cal B}^A = (\delta^A,{}^A\mu_t^{ij})$, surplus is generated at rate $W_t^i$ at time $t$, during a sojourn in state $i$, as follows. (Note we allow $\delta^L \not=\delta^A$ here just for the purposes of demonstration.) Under Thiele's equation applied to $\tilde{\cal B}^L$:

\begin{equation}
\frac{d}{dt} V_t^i = \delta^L \, V_t^i + \pi^L_i - \sum_{j \not=i} {}^L\mu_t^{ij} \, R_t^{ij} \label{eq:ThieleL}
\end{equation}

\noindent where $V_t^i$ is the policy value, $\pi_i^L$ is the valuation premium rate and $R_t^{ij} = (b_{ij} + V_t^j - V_t^i)$, see Section \ref{sec:ThieleMarkov}. Applying the parameters of ${\cal B}^A$ to $V_t^i$ we have:

\begin{equation}
\frac{d \, V_t^i}{dt} + W_t^i = \delta^A \, V_t^i + P_i - \sum_{j \not=i} {}^A\mu_t^{ij} \, R_t^{ij} \label{eq:ThieleA}
\end{equation}

\noindent (recall that we assume all accumulation bases operate on the contractual premium rates $P_i$) and by subtraction of equation (\ref{eq:ThieleL}) from (\ref{eq:ThieleA}):

\begin{equation}
W_t^i = (\delta^A - \delta^L) \, V_t^i + (P_i - \pi_i^L) - \sum_{j \not=i} ( {}^A\mu_t^{ij} - {}^L\mu_t^{ij}) \, R_t^{ij}. \label{eq:ThieleDiff}
\end{equation}

\noindent (From this point we resume the assumption of constant $\delta$, hence {there is} no contribution to surplus from investment returns.) Thus the present value of total surplus up to time $t$, allowing for any generated at inception can be written as:

\begin{equation}
\underbrace{\int_0^t v^s \, W_s^{J(s)} \, ds}_{\mbox{\scriptsize State Space representation}} - \quad V_0^1 = \underbrace{\sum_i \int_0^t v^s \, Y_s^i \, W_s^i \, ds}_{\mbox{\scriptsize Counting Process representation}} - \quad V_0^1. \label{eq:AccSurpluses}
\end{equation}

We make some comments below on the functions $W_t^i$.

\begin{bajlist}

\item In any multiple-decrement model with one `live' state $i=1$, the function $W_t^1$ is identical to the `critical function' of Lidstone's theorem \citep{lidstone1905}, see \cite{norberg1985}.

\item {Given valuation basis ${\cal B}^L$ and accumulation basis ${\cal B}^A$, define the EPV of surplus to time $t$ to be $\Gamma_t^{L,A}$ as follows:

\begin{equation}
\Gamma_t^{L,A} = \int_0^t v^s \, W_s^{J(s)} \, ds = \sum_i \int_0^t v^s \, Y_s^i \, W_s^i \, ds
\end{equation}

\noindent as in equation (\ref{eq:AccSurpluses}), with surplus rates $W_t^i$ defined by ${\cal B}^L$ and ${\cal B}^A$.}

\item {The functions $W_t^i$ do not depend on any boundary conditions of Thiele's equations, just the parameters. Therefore `surplus' during the policy term is just a relative quantity that can be applied to measure the difference between any pair of technical bases of any type, not confined to an accumulation basis and a valuation basis.}

\end{bajlist}


\subsection{Surplus and the Counting Process Representation of the Data}
\label{sec:RamlauHansen1988}


In a fundamental paper, \cite{ramlau-hansen1988a} defined the present value of surplus based on the counting process $N_t = \{N_t^{ij}\}$. We have changed his notation to be consistent with ours. He assumed a constant force of interest $\delta$ and the same statewise premium rates $P_i$ under all technical bases, therefore the intensities were the only sources of surplus. He also assumed $V_0^1=0$. The present value of surplus to time $t$, denoted by $\Gamma_t^L$, is the random variable:

\begin{equation}
\Gamma_t^L = \sum_i \int_0^t v^s \, Y_s^i \, P_i \, ds - \sum_i \sum_{j \not= i} \int_0^t v^s \, b_{ij} \, dN^{ij}_s - v^t \, V_t^{J(t)} \label{eq:Gamma}
\end{equation}

\noindent {in which policy values $V_t^i$ are on technical basis ${\cal B}^L.$} {Note that the first part of this expression is obtained from the first term in equation (\ref{eq:Split1}) upon defining:

\begin{equation}
A(t) = \int_0^t \sum_i \left( Y_s^i \, P_i \, ds - \sum_{j \not=i} b_{ij} \, dN_s^{ij}  \right).
\end{equation}

{The processes $\Gamma_t^{L,A}$ and $\Gamma_t^L$ are both stochastic processes, the first defined by the difference between technical bases ${\cal B}^A$ and ${\cal B}^L$, the second defined by actual cashflows and technical basis ${\cal B}^L$. Note in particular that $\Gamma_t^{L,M} \not= \Gamma_t^L$.}

\cite{ramlau-hansen1988a} gave two expressions for $\Gamma_t^L$ in terms of model quantities $\{{}^{M}\mu_t^{ij}\}$ and $\{{}^L\mu_t^{ij}\}$ and a martingale residual. The first, modified here to allow for $V_0^1 \not= 0$, was in terms of accumulated surplus:

\begin{eqnarray}
\Gamma_t^L & = & \sum_i \sum_{j \not= i} \int_0^t v^s \, Y_s^i \, ( {}^L\mu_s^{ij} - {}^M \! \mu_s^{ij} ) \, R_s^{ij} \, ds - \sum_i \sum_{j \not= i} \int_0^t v^s \, R_s^{ij} \, dM_s^{ij} - V_0^1. \label{eq:Gamma2}
\end{eqnarray}

\begin{bajlist}

\item \cite{wolthuis1990} remove the assumption that both  technical bases have the same force of interest $\delta_t$. Then the rate of surplus earned includes a term $({}^A\delta_t - {}^L\delta_t) \, V_t^{J(t)}$.

\item We retain the assumption that both  technical bases have the same constant force of interest $\delta$, just to simplify expressions. We do not assume that premium rates are the same on each technical basis, we may have $P_i \not= \pi_i^L \not= \pi^*({\cal B}^L)$.

\end{bajlist}

The second expression for $\Gamma_t^L$ was in terms of an accumulated fund, which we call $A^i_t$. In \cite{ramlau-hansen1988a} this was a retrospective policy value on the second-order technical basis, with initial value zero, hence consistent with our definition of accumulation fund. Define ${}^A \! R_t^{ij} = b_{ij} + A_t^j - A_t^i$. Then:

\begin{equation}
\Gamma_t^L = v^t \, (A_t^{J(t)} - V_t^{J(t)}) - \sum_i \sum_{j \not= i} \int_0^t v^s \, {}^A \! R_s^{ij} \, dM_s^{ij} \label{eq:Gamma3}
\end{equation}

\noindent (and here $\Gamma_0^L = -V^1_0$ is not necessarily zero). The author comments: ``The relation (\ref{eq:Gamma3}) is interesting because it yields an expression for $\Gamma_t^L$ in terms of $v^t \, (A_t^{J(t)} - V_t^{J(t)})$, the mean of which is often interpreted as the gain obtained over $[0,t]$ when a deterministic approach to life contingencies is applied'' (notation has been changed to agree with ours).

The following result (\cite[(4.3)]{ramlau-hansen1988b} with $V_0^1$ not necessarily zero) will be needed later:

\begin{equation}
\sum_i \int_0^t Y_s^i \, \frac{d \, (v^s \, V_s^i)}{ds} \, ds + \sum_i \int_0^t v^s \sum_{j \not= i} (V_s^j - V_s^i) \, dN_s^{ij} = v^t \, V_t^{J(t)} - V_0^1. \label{eq:RHKeyResult}
\end{equation}

\noindent It allows us to integrate functions along sample paths $J(t)$ allowing for jumps, if we know their statewise values, in this case $V_t^i$.

\newpage

\subsection{Relationships Between Technical Bases in Multiple-state Models}
\label{sec:RelBasesMarkov}

We wish to demonstrate the relationships {\bf R1} to {\bf R3} between pairs of technical bases, as stated in Section \ref{sec:Motivation}. Moreover, we would like to do so for  modelled rates of surplus $W_t^i$ based on an arbitrary accumulation basis ${\cal B}^A$. {The quantities we deal with, including {$W_t^{J(t)}$}, are stochastic, that is ${\cal F}_t$-measurable, and expectations are  conditional expectations given ${\cal F}_0$.}


Relationship {\bf R1} between premium and valuation bases is straightforward (Section \ref{sec:InitialSurp}) since it is independent of any accumulation basis. In Section \ref{sec:SurplusTermMarkov}, which largely follows \cite{ramlau-hansen1988a}, we integrate statewise rates of earned surplus $W_t^i$ along a sample path to find relationship {\bf R2}. Finally in Section \ref{sec:FinalProfitMarkov} we find relationship {\bf R3} but show that the final profit is independent of the valuation basis only if the accumulation basis is ${\cal B}^M$.


\subsubsection{Relationship {\bf R1}: Initial Surplus: Premium and Valuation Bases}
\label{sec:InitialSurp}

\begin{proposition} Let ${\cal B}^L = (\delta^L,{}^L\mu_t^{ij})$, and define $\pi^* = \pi^*({\cal B}^L)$, the pure premium rates associated with ${\cal B}^L$, and $\pi^L = \pi(\tilde{\cal B}^L)$, the valuation premium rates associated with $\tilde{\cal B}^L$. Then:

\begin{equation}
V_0^1 = \int_0^{n} v^s \, \sum_{j=1}^m P_{1j}(0,s) \,  (\pi^*_j - \pi_j^L) \, ds, \label{eq:CapLoadingsMarkov}
\end{equation}

\noindent {or in words,} equal to (minus) the present value of premium loadings capitalized at outset. \end{proposition}

\noindent {\em Proof}: Write initial surplus at time $t=0$ as:

\begin{equation}
V_0^1 = \int_0^{n} v^s \, \sum_{j=1}^m P_{1j}(0,s) \, \bigg[ - \pi_j^L + \sum_{k \not= j} {}^L\mu_{s}^{jk} \, b_{jk} \bigg] \, ds + v^n \, \sum_{j=1}^m P_{1j}(0,n) \, M_j.
\end{equation}

\noindent Add and subtract $\int_0^n v^s \, \sum_{j=1}^m P_{1j}(0,s) \, \pi^*_j$, and from the definition of $\pi^*(\tilde{\cal B}^L)$, the result follows. \hfill{$\Box$}

\medskip

{Corollary 1 is slightly out of logical sequence here, since its proof calls upon Proposition 2, but we take it where it fits most naturally.}

\begin{corollary} Each contractual premium rate $P_i$ can be decomposed into:

\begin{equation}
P_i = \pi^*_i + (\pi^L_i - \pi^*_i) + (P_i - \pi^L_i) \label{eq:Decomp}
\end{equation}

\noindent in which the terms on the right-hand side are, respectively, a pure risk premium, a loading capitalized at outset and a loading falling into surplus as premiums are paid.
\end{corollary}

\noindent (Compare with \cite{linnemann2003}, who bases loadings on a second-order basis equating to the experience basis.)
The regulator may apply external constraints to ensure that $P_i \ge \pi_i^L$ and $\pi^L_i \ge \pi^*_i$; we do not.

{\noindent {\em Proof of Corollary 1}: It is obvious that $P_i$ can be decomposed as in equation (\ref{eq:Decomp}), and $\pi^*_i$ is the pure risk premium in state $i$ on technical basis ${\cal B}^L$ by definition. Proposition 1 shows that a loading of $\pi_i^L - \pi^*_i$ is capitalized at outset. Proposition 2 will show that the loading $P_i - \pi_i^L$ appears in the surplus at time $t$ $(0 < t < n)$. \hfill{$\Box$}.

\begin{bajlist}

\item For example, if the premium basis is ${\cal B}^P = (\delta,{}^P\mu_t^{ij})$ and $\tilde{\cal B}^L$ is a gross premium valuation basis $((\delta,{}^L\mu^{ij}_t),\pi_i^L)$, then $(\pi_i^* - \pi^L_i) = (\pi^*_i - P_i)$ so $V^1_{0}$ is (minus) the EPV of the premium loadings $(P_i - \pi^*_i)$, which therefore fall into surplus at outset.

\item Or, given a net premium valuation basis $\tilde{\cal B}^L = ((\delta,{}^L\mu^{ij}_t),\pi^*_i)$, the loading $(\pi^*_i - \pi_i^L) = 0$, so there is no initial surplus, and all loadings $(P_i - \pi^L_i)$ fall into surplus as premiums are paid {(see Proposition 2)}.  

\item {While (a) and (b) above probably cover the major examples of interest, other possibilities exist. An example was given in Section \ref{sec:DefsTBandPP}. Or, it would be possible to fix the valuation premium rates $\pi^L$ such that $(P_i - \pi^L_i) = (\pi^L_i - \pi^*_i) = (P_i - \pi^*_i)/2$, so that half the premium loadings are capitalized at outset and half are not.}

\end{bajlist}

Everything said above, qualitative and quantitative, is independent of the accumulation basis ${\cal B}^A$, which is intuitively obvious because at time $t=0$ there has not yet been any experience.


\subsubsection{Relationship {\bf R2}, Surplus During the Term: Valuation and Accumulation Bases}
\label{sec:SurplusTermMarkov}

Recall that our purpose here is to express relationship {\bf R2} for modelled surpluses $\Gamma_t^{L,A}$ given by the rates $W_t^i$; Ramlau-Hansen's equation (\ref{eq:Gamma2}) did so for crude surplus $\Gamma_t^L$. 

\begin{proposition} {(following Ramlau-Hansen).} For time $t$, $0 < t < n$, and defining $\Delta {}^A\mu_t^{ij} = {}^A\mu_t^{ij} - {}^{M}\mu_t^{ij}$, the present value of total surplus including initial surplus is:

\begin{eqnarray}
\int_0^t v^s \, W_s^{J(s)} \, ds - V_0^1 & = & \sum_i \int_0^t v^s \, \left[ Y_s^i \, P_i \, ds - \sum_{j \not= i} b_{ij} \, dN_s^{ij} \right] - v^t \, V_t^{J(t)} \nonumber \\
&   & \quad - \sum_i \int_0^t v^s \, Y_s^i \, \sum_{j \not= i} \Delta {}^{A}\mu_s^{ij} \, R_s^{ij} \, ds + \sum_i \int_0^t v^s \sum_{j \not= i} R_s^{ij} \, dM_s^{ij} \nonumber \\
&   & {} \nonumber
\end{eqnarray}

\noindent which can be written as:

\begin{eqnarray}
\Gamma_t^{L,A} - V_0^1 & = & \Gamma_t^L - \sum_i \int_0^t v^s \, Y_s^i \, \sum_{j \not= i} \Delta {}^{A}\mu_s^{ij} \, R_s^{ij} \, ds + \sum_i \int_0^t v^s \sum_{j \not= i} R_s^{ij} \, dM_s^{ij}. \\
&   & {} \nonumber
\end{eqnarray}

\end{proposition} 

\noindent {\em Proof}: Consider the derivative of $v^t \, V_t^i$:

\begin{eqnarray}
\frac{d}{dt} \, ( v^t \, V_t^i) & = & - \delta \, v^t \, V_t^i + v^t \, \left[ \delta \, V_t^i + \pi_i^L - \sum_{j \not= i} {}^L\mu_t^{ij} \, R_t^{ij} \right].
\end{eqnarray}

\noindent Add and subtract $v^t \, ( P_i - \sum_{j \not= i} {}^A\mu_t^{ij} \, R_t^{ij} )$ to/from the right-hand side:

\begin{eqnarray}
\frac{d}{dt} \, ( v^t \, V_t^i) & = & - v^t \, \left[ ( P_i - \pi_i^L ) - \sum_{j \not=i} ({}^A\mu_t^{ij} - {}^L\mu_t^{ij}) \, R_t^{ij} \right] \nonumber \\
&   & \quad + \,\, v^t \, \left[ P_i - \sum_{j \not= i} {}^A\mu_t^{ij} \, R_t^{ij} \right] \nonumber \\
& = & - v^t \, W_t^i + v^t \, \left[ P_i - \sum_{j \not= i} {}^A\mu_t^{ij} \, b_{ij} \right] - v^t \, \sum_{j \not= i} {}^A\mu_t^{ij} \, (V_t^j - V_t^i). \label{eq:SurpExpression}
\end{eqnarray}

\noindent Therefore, in terms of rates during sojourns in state $i$:

\begin{eqnarray}
v^t \, W_t^i & = & v^t \, \left[ P_i - \sum_{j \not= i} {}^A\mu_t^{ij} \, b_{ij} \right] - \left[ \frac{d \, (v^t \, V_t^i)}{dt} + v^t \, \sum_{j \not= i} {}^A\mu_t^{ij} \, (V_t^j - V_t^i) \right]. \label{eq:PVSurpAccumBasis}
\end{eqnarray}

\noindent We then proceed {\em via} the following steps, omitting some extensive but elementary substitutions at (b): (a) multiply throughout by $Y_t^i$; (b) add/subtract $v^t \, \sum_{j \not= i} R_t^{ij} \, dM_t^{ij}$ to/from the right-hand side, and rearrange, defining $\Delta {}^A\mu_t^{ij} = {}^A\mu_t^{ij} - {}^{M}\mu_t^{ij}$; and (c) integrate on $[0,t]$ and sum over all states $i$, to obtain:

\begin{eqnarray}
\sum_i \int_0^t v^s \, Y_s^i \, W_s^i \, ds & = & \sum_i \int_0^t v^s \,  \left[ Y_s^i \, P_i \, ds - \sum_{j \not= i} b_{ij} \, dN_s^{ij}  \right] \nonumber \\
&   & \quad - \sum_i \int_0^t \left[ Y_s^i \, \frac{d \, (v^s \, V_s^i)}{ds} \, ds + v^s \, \sum_{j \not= i} (V_s^j - V_s^i) \, dN_s^{ij} \right] \label{eq:PVSurpAccumBasis5} \nonumber \\
&   & \quad + \sum_i \int_0^t v^s \sum_{j \not= i} R_s^{ij} \, dM_s^{ij} - \sum_i \int_0^t v^s \, Y_s^i \sum_{j \not= i} \Delta {}^A\mu_s^{ij} \, R_s^{ij} \, ds. \nonumber \\
& & \label{eq:TotalSurp3}
\end{eqnarray}

\noindent Now apply the result in equation (\ref{eq:RHKeyResult}) to integrate the middle line above:

\begin{eqnarray}
\underbrace{\sum_i \int_0^t v^s \, Y_s^i \, W_s^i \, ds - V_0^1}_{\mbox{\scriptsize PV[Total Surplus]}} & = & \underbrace{\sum_i \int_0^t v^s \, \left[ Y_s^i \, P_i \, ds - \sum_{j \not= i} b_{ij} \, dN_s^{ij} \right] - v^t \, V_t^{J(t)}}_{\mbox{\scriptsize $\Gamma_t^L$ = PV[Assets $-$ Liabilities on ${\cal B}^L$]}} \nonumber \\
&   & \quad - \underbrace{\sum_i \int_0^t v^s \, Y_s^i \, \sum_{j \not= i} \Delta {}^{A}\mu_s^{ij} \, R_s^{ij} \, ds}_{\mbox{\scriptsize Systematic Surplus}} + \underbrace{\sum_i \int_0^t v^s \sum_{j \not= i} R_s^{ij} \, dM_s^{ij}}_{\mbox{\scriptsize Martingale}}. \nonumber \\
&   & {} \label{eq:AccumSurpFinal}
\end{eqnarray}

\hfill{$\Box$}

\begin{corollary} {\noindent The EPV $\Gamma_t^L$ of the cumulative surplus at any time $t$ $(0 < t < n)$ can be decomposed into: (a) initial surplus, plus; (b) premium loadings, plus; (c) a sum of pairs, each pair consisting of a systematic part and a martingale residual, for each source of surplus. Moreover, $\Gamma_t^L$ does not depend on the premium basis ${\cal B}^P$, once the contractual terms have been fixed.}

\end{corollary}

\noindent {\em Proof}: Recall that equation (\ref{eq:Gamma2}) excluded loading surplus, {but the derivation of equation (\ref{eq:AccumSurpFinal}) included it}. If we { include loading surplus at rate $(P_i - \pi^L_i)$ in equation (\ref{eq:Gamma2})}, and substitute the result into equation (\ref{eq:AccumSurpFinal}) we have:

\begin{eqnarray}
\sum_i \int_0^t v^s \, Y_s^i \, W_s^i \, ds & = &  \sum_i \int_0^t v^s \, Y_s^i \, \left[ (P_i - \pi^L_i) - \sum_{j \not= i} ( {}^{A} \! \mu_s^{ij} - {}^L \! \mu_s^{ij} ) \, R_s^{ij} \, ds \right]  \label{eq:AccumSurpFinal2}
\end{eqnarray}

\noindent {and substituting this in equation (\ref{eq:AccumSurpFinal}) and re-arranging gives the result. It is plain upon inspection that the right-hand side of equation (\ref{eq:AccumSurpFinal}) does not depend on ${\cal B}^P$ once the contractual basis is fixed. Equation (\ref{eq:AccumSurpFinal2}) shows that the same is true of the left-hand side. 

\hfill{$\Box$}}

Comparing equations (\ref{eq:ThieleDiff}) and (\ref{eq:AccumSurpFinal2}), we see the latter simply displays some operations performed on the former, in the absence of interest surplus, so offers another possible derivation of equation (\ref{eq:AccumSurpFinal}). However it is obtained, we need the representation in equation (\ref{eq:AccumSurpFinal}) for relationship {\bf R3}, see Section \ref{sec:FinalProfitMarkov}.

{The term `systematic surplus' for the penultimate term in equation (\ref{eq:AccumSurpFinal}) is from \cite[p.213]{moeller2007}, see Section \ref{sec:OtherExample}. These authors suggest two ways to smooth the random component of total surplus from equation (\ref{eq:AccumSurpFinal}): (i) take expectations conditioning on ${\cal F}_0$ as in \cite{norberg1991} (see Section \ref{sec:StochDecomp}); or (ii) ignore both systematic and martingale components of the surplus. \cite{schilling2020} suggest a decomposition of total surplus (including investment surplus) into systematic and (orthogonal) martingale components of which equation (\ref{eq:AccumSurpFinal}) shows those terms due to the transition intensities. \cite{jetses2022} develop this as the `infinitesimal sequential updates' (ISU) decomposition principle for surplus, avoiding difficulties associated with finite accounting periods.}


\subsubsection{Relationship {\bf R3}, Profit: Premium and Accumulation Bases}
\label{sec:FinalProfitMarkov}

\begin{proposition} The EPV of total final surplus $\Mean[\Gamma_n^{L,A}] - V_0^1$ is $\Mean[\Gamma_n^L]$ minus the EPV of the difference in risk premiums arising from any difference between ${\cal B}^A$ and ${\cal B}^M$, plus a martingale residual term. \end{proposition}

\noindent {\em Proof}: Evaluate equation (\ref{eq:AccumSurpFinal2}) at $t=n$ and take expectations. \hfill{$\Box$}

\begin{corollary} \noindent If in Proposition 3, ${\cal B}^A = {\cal B}^{M}$, then the EPV of total surplus is independent of the valuation basis. 

\end{corollary}

\noindent {\em Proof}: If ${\cal B}^A = {\cal B}^{M}$, then in the right-hand side of equation (\ref{eq:AccumSurpFinal}) evaluated at $t=n$ all $\Delta {}^{A}\mu_s^{ij} = 0$, and policy values are involved only through $\Mean[v^n \, V_n^{J(n)}]$, which does not depend on the valuation basis. \hfill{$\Box$}

If ${\cal B}^A \not= {\cal B}^{M}$ then the EPV of the total surplus is not independent of the valuation basis, because of the presence of the $R_t^{ij}$ in the systematic surplus.

\subsection{Comment on Introducing Technical Basis ${\cal B}^M$}
\label{sec:CommentIntroCP}

It is tempting to interpret the last term (in large square brackets) in equation (\ref{eq:PVSurpAccumBasis}) as follows: (a) the derivative gives us the change in $v^t \, V_t^i$ between jumps, which we can integrate as usual; and (b) the second term (times $dt$) is the expected change in $v^t \, V_t^i$ on a jump to state $j$ at time $t$, so we can obtain $\Mean[ \, \sum_i \int_0^t Y_s^i \, v^s \, W_s^i \, ds \, ]$ by applying the same operations term-by-term to the right-hand side of equation (\ref{eq:PVSurpAccumBasis}).

{However, as noted in Section \ref{sec:ExpectedValues}, all expected values are assumed to be with respect to the `true' technical basis ${\cal B}^M$ and} we are not free to interpret terms in ${}^A\mu_t^{ij}$ (times $dt$) as expectations. Hence in the step between equations (\ref{eq:PVSurpAccumBasis}) and (\ref{eq:TotalSurp3}) we have to swap out the terms in ${}^A\mu_t^{ij}$ for terms in ${}^M \! \mu_t^{ij}$ in order to apply equation (\ref{eq:RHKeyResult}).


\section{Examples}
\label{sec:Examples}

\subsection{Example 1: Lapse-supported Business}
\label{sec:LSB}

Lapse-supported business is a class of non-participating business mainly written in North America. The policies are whole-of-life, or endowments to a high age, depending on local practice\footnote{Valuation regulations may require a published life table to be used. If published tables cease at age 100 (for example) this dictates the design of the contract. In Canada the main contract is an endowment ceasing at age 100, confusingly called `Term to 100'.}; we assume they are endowments to age 100, with policy term $n = 100-x$. Surrender values are as small as possible, so ignoring expenses, and assuming policy values are positive, we assume that lapses are profitable at all durations. Therefore, premium rates can be reduced by allowing for future lapses. This forms the basis of a competitive market, particularly in Canada. However, insurers are exposed to the risk of lapse rates being lower than anticipated. See \cite{hacariz2023} for details.


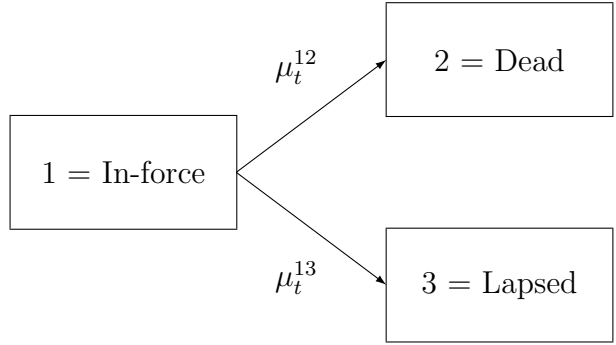
\begin{figure}
\begin{center}
\begin{picture}(95,50)
\put(50,30){\framebox(30,15){}}
\put(50,0){\framebox(30,15){}}
\put(0,15){\framebox(30,15){}}
\put(30,22.5){\vector(4,-3){20}}
\put(30,22.5){\vector(4,3){20}}
\put(15,22.5){\makebox(0,0)[c]{1 = In-force}}
\put(65,7.5){\makebox(0,0)[c]{3 = Lapsed}}
\put(65,37.5){\makebox(0,0)[c]{2 = Dead}}
\put(38,36.5){\makebox(0,0)[c]{$\mu_{t}^{12}$}}
\put(38,8.5){\makebox(0,0)[c]{$\mu_{t}^{13}$}}
\end{picture}
\end{center}
\caption{\label{fig:Lapses}A Markov model of transfers between states representing in-force life insurance, death and lapsation.}
\end{figure}

\begin{bajlist}

\item {\em Model}: The underlying model may be represented as in Figure \ref{fig:Lapses}, in which the age at issue $x$ is suppressed as usual. 

\item {\em Contract}: Suppose the lapse-supported contract has premium rate $P$ per year, death benefit $S$, maturity benefit $M$ at age 100, and anticipates a surrender value of $C_t$ at duration $t$. This represents an extension of the model to a time-varying benefit $C_t$, which we allow without further comment. When considering alternative technical bases, the rule that level premiums are paid while in state 1, and none in states 2 or 3, will ensure that the principle of equivalence can be applied in all sensible cases to calculate pure premiums $\pi^*({\cal B})$.

\item {\em Technical Bases}: A technical basis may be denoted by ${\cal B} = (\delta, \mu_t^{12} , \mu_t^{13})$, with associated pure premium $\pi^* = \pi^*({\cal B})$ since premiums are payable in state 1 only. Assume that all technical bases have the same interest and mortality, so differ only in respect of lapse intensities and premium rates. Our simplest baseline setup is that the premium, valuation and accumulation bases are the same, echoing the model in Section \ref{sec:ModelFund}; moreover the accumulation basis is the `true' ${\cal B}^M$: 

\begin{equation}
{\cal B}^{P} = {\cal B}^L = {\cal B}^A = {\cal B}^M = (\delta , \mu_t^{12} , \mu_t^{13}) \label{eq:LSBBasisP}
\end{equation}

\noindent and {$\tilde{\cal B}^L = ({\cal B}^L , P)$}. Recall that $P=\pi^*({\cal B}^P)$ by definition. Define an alternative valuation basis $\tilde{\cal B}^{L^*}$ as follows:

\begin{equation}
{\cal B}^{L^*} = (\delta , \mu_t^{12} , 0 ) \qquad \mbox{and} \qquad \tilde{\cal B}^{L^*} = ({\cal B}^{L^*} , P^*) \label{eq:LSBBasisL*}
\end{equation} 

\noindent where $P^* = \pi^*({\cal B}^{L^*})$. Technical basis $\tilde{\cal B}^{L^*}$ is a proper valuation basis {\em for the same contract ${\cal C}$} assuming nil lapses and with pure premium $P^*$.

\item {\em Relationship {\bf R1}}: The contractual premium rate $P$ decomposes into risk premium plus loading as follows: $P = P^* + (P - P^*)$. Since $P < P^*$ the loading is negative. 

\item {\em Relationship {\bf R3}}: The EPV of total surplus, where expectations are taken under ${\cal B}^M$, does not depend on the valuation basis, ${\cal B}^L$ or ${\cal B}^{L^*}$. Both are proper valuation bases so ${}^L \! V_0^1 = {}^{L^*} \! V_0^1 = 0$. It is straightforward to calculate rates of surplus emerging during stays in state 1 at times $t \ge 0$, as:

\begin{equation}
{}^L \! W_t^1 = - ( {}^M \! \mu_t^{13} - {}^L \! \mu_t^{13} ) (C_t - {}^L \! V_t^1)
\end{equation}

\noindent and:

\begin{equation}
{}^{L^*} \! W_t^1 = (P - P^*) - {}^M \! \mu_t^{13} \, (C_t - {}^{L^*} \! V_t^1)
\end{equation}

\noindent respectively. Hence, defining $\psi_t = \exp ( - \int_0^t ( \delta + \mu_s^{12} + {}^M \! \mu_s^{13} ) \, ds ) $ to be the discount factor allowing for survivorship under ${\cal B}^M$, integrating and re-arranging, the EPV of future premiums `mortgaged' by assuming lapse-support is:

\begin{equation}
\int_0^{\infty} \psi_s \, ( P - P^* ) \, ds = \int_0^{\infty} \psi_s \, {}^L \! \mu_s^{13} \, (C_s - V^*_s) \, ds + \int_0^{\infty} \psi_s \, {}^M \! \mu_s^{13} \, (V^*_s - V_s) \, ds \label{eq:LSBDiff1}
\end{equation}

\noindent (see \cite{hacariz2023}). The first term on the right-hand side is the EPV of the lapse surpluses anticipated, while the second term adjusts for the shortfalls on actual lapses. If all $C_t=0$ and ${}^L \! \mu_t^{13} = {}^M \! \mu_t^{13}$, which means that lapses and lapse surplus are anticipated to the maximum possible extent, then:

\begin{equation}
\int_0^{\infty} \psi_s \, (P - P^*) \, ds = - \int_0^{\infty} \psi_s \, {}^M\! \mu_s^{13} \, V_s \, ds \label{eq:LSBDiff2}
\end{equation}

\noindent which is the EPV of policy values on all lapsed policies. 

\end{bajlist}

The quantities in equations (\ref{eq:LSBDiff1}) and (\ref{eq:LSBDiff2}) are negative, therefore they measure a  need for capital support, rather than showing surplus being released. This aspect of lapse-supported business is, of course, well-known in practice, see \cite{record1987}, and may well lead to there being reserving requirements {\em a posteriori}. 

{We could turn this example around and define ${\cal B}^{P} = {\cal B}^A = {\cal B}^M = (\delta , \mu_t^{12} , 0 )$, and put $\tilde{\cal B}^{L} = ({\cal B}^{L},\pi^*({\cal B}^{L}))$, so the premium and valuation bases allow for zero lapses, and then put ${\cal B}^{L^*} = (\delta , \mu_t^{12} , \mu_t^{13} )$ and $\tilde{\cal B}^{L^*} = ({\cal B}^{L^*},P^*)$ where $P^* = \pi^*({\cal B}^{L^*}))$, and then $\tilde{\cal B}^{L^*}$ is an alternative proper valuation basis allowing for lapses, for the same contract.} Then the question would be whether ${\cal B}^{L^*}$ was sufficiently prudent to allow loadings $(P-P^*)$ to be capitalized. It would be interesting to know if the regulator's response to this question depended on which way  it was framed. Either way, this example demonstrates freedom from {\em a priori} constraints on technical bases.


\subsection{Example 2: \cite{moeller2007}}
\label{sec:OtherExample}

\cite{moeller2007} define a model life insurer's balance sheet in which prospective policy values have been subordinated. The aim is to admit retrospective considerations to the definition of the office's  liabilities, {{\em via} the conversion of recognized surplus to bonus}. The authors consider a contract with term $n$ years, annual premium rate $\pi$, death benefit $b^{ad}$, and maturity benefit $b^a(t)$. They define three technical bases, differing from those in Section \ref{sec:ModelBalanceSheet}, as follows.

\begin{bajlist}

\item A {\em first-order technical basis} denoted by $(r^*,\mu^*)$ which at time $t$ defines the maturity benefit $b^{a}(t)$ given by the premium rate $\pi$ and technical reserve $V^*(t)$ (see (b) below) under the equivalence principle. Thus, at time $t=0$, this technical basis performs the traditional r\^ole of premium basis. At times $t > 0$ it determines an entitlement to an increased maturity benefit $b^a(t) > b^a(0)$, therefore a bonus.

\item A {\em second-order technical basis} denoted by $(r^{\delta},\mu^{\delta})$ which defines the {\em technical reserve} $V^*(t)$ under Thiele's equation, with initial boundary condition $V^*(0)=0$.

\item A {\em third-order technical basis} denoted by $(r,\mu)$ which defines the {\em total reserve} $U(t)$ under Thiele's 
equation, with initial boundary condition $U(0)=0$.

\end{bajlist}

\noindent All three technical bases are associated with the same premium rate $\pi$. The general idea, expressed in our notation, is of  a `safe-side' premium and valuation basis {${\cal B}^P = (r^*,\mu^*)$; a policy account $V^*(t)$ based on a somewhat less conservative accumulation basis ${\cal B}^A = (r^{\delta},\mu^{\delta})$; and the `true' accumulation basis ${\cal B}^M = (r,\mu)$} under which the asset share $U(t)$ builds up. 

There is, in addition, a {\em valuation basis} that we call ${\cal B}^L$, with policy values denoted by $V(t)$, defined by the `real' technical basis $(r,\mu)$, valuation premium rate $\pi$ and and terminal boundary condition $V(n) = b^a(n)$. Note that this boundary condition is not a contractual benefit but a solution of Thiele's equation under technical basis {${\cal B}^P$} above. Such a possibility exists in the quasi-stochastic setup {of traditional life insurance mathematics}, in which all objects are conditional expectations with respect to ${\cal F}_0$; it is not obvious that it exists in a model in which $b^a(n)$ must be acknowledged to be {only} ${\cal F}_n$-measureable.

The {\em undistributed reserve} $X(t)$ at time $t$ is defined by $X(t) = U(t) - V^*(t)$, it being supposed that $V^*(t)$ has been distributed {by time $t$}, for example by being allocated to a policy account, whether or not that fact has been disclosed. A key assumption is that $X(n) = U(n) - V^*(n) = 0$ (p.16), which simplifies the model greatly. Some consequences of this assumption are as follows.

\begin{bajlist}

\item It is then evident that $V(t) = U(t)$ for all $t \ge 0$.

\item The setup simplifies so that, apart from the premium basis ${\cal B}^P$, we have just two accumulation bases ${\cal B}^A$ and ${\cal B}^M$ which have common boundary values 0 and $b^a(n)$ (meaning, incidentally, that they satisfy the sufficient conditions of Lidstone's theorem (\cite{lidstone1905}, \cite{norberg1985})).

\item The second-order basis drops out of certain calculations involving total profit (``$\ldots$ further specification of the future second order basis is redundant'', p.17). In our terms, this is Corollary 3 (Section \ref{sec:FinalProfitMarkov}), the irrelevance of the valuation basis. It is stated that ``$\ldots$ the condition $X(n) = 0$ is the same as performing the equivalence principle on the total payments under the real basis'' (p.17).
 
\item For the purposes of demonstration only, suppose that interest is the only source of surplus. Then assuming $X(n)=0$, equation (2.15) of \cite{moeller2007} reduces to:

\begin{equation}
\int_0^n e^{ - \int_0^t (r(s) + \mu(s)) \, ds} \,(r(t) - r^{\delta}(t)) \, V^*(t) \, dt = 0. 
\end{equation}

\noindent Ruling out other possibilities, either $r(s) = r^{\delta}(s)$ on $[0,n]$, or $(r(s) - r^{\delta}(s))$ must change sign at least once on $[0,n]$. So the setup cannot simply be that second-order ${\cal B}^A$ is a uniformly `weaker' accumulation basis than third-order ${\cal B}^M$.

\end{bajlist}

It is noted that when conditional expectations are encountered based on first-order $\mu^*$ and second-order $\mu^{\delta}$, then ``$\ldots$ since the intensity of $N$ in the conditional expectation is not $\mu$, these quantities can only be said to build on suitable imitations of the principles'', (p.27) (compare with Section \ref{sec:CommentIntroCP}, which introduces the privileged position of technical basis ${\cal B}^M$ and its intensities). 

A final feature of the model is that, in the absence of a true terminal bonus system, for example as it is known in the UK, all bonus that will be distributed must be included in $V^*(n)$ by the end of the term, which must be done by choosing the technical basis ${\cal B}^A$ to hit the target $X(n)=0$. It is said that ``The second order basis is a decision variable held by the insurer that is to be chosen within certain legislative constraints and market conditions'' (p.13). So the actuary still faces the classical challenge of using only reversionary bonuses to hit an asset share target.


\section{Conclusions}
\label{sec:Concs}

Almost all of the literature on Thiele's equation and surplus in life insurance uses a model {inspired by Scandinavian-style regulation} with two technical bases, called first-order and second-order. This is more restrictive than practice in some jurisdictions, {excludes common valuation methodologies}, and uses terminology that may not be universally familiar. {The literature} emphasizes retrospective policy values, which are uniformly familiar but, arguably, obsolete.

Our setting is Markov models, defined in Section \ref{sec:TechBase}. Given a particular contract, we propose: (a) a definition of `technical basis' ${\cal B}$ that includes interest and transition intensities (expenses and anything else can be added if desired); {(b) the `pure' premium rate $\pi^*({\cal B})$ resulting if we plug technical basis ${\cal B}$ into the principle of equivalence; and (c) for technical bases classified as valuation bases (see below) the `valuation' premium rate $\pi({\cal B})$.}

We then classify technical bases based on the boundary conditions that are satisfied in Thiele's equation (Section \ref{sec:ClassificationOverall}). A `valuation basis' satisfies the terminal boundary conditions $V_n^i=M_i$ $(i \in {\cal S})$ and defines policy values. An `accumulation basis' satisfies the initial boundary {condition $V_0^1=0$ (assuming everyone starts in state 1)} and defines a policy account. We have no {need of} the traditional retrospective policy value.

{We suppose there is a `true' accumulation basis denoted by ${\cal B}^M$ and called the {experience} basis, which has a privileged position. Its transition intensities ${}^M \! \mu_t^{ij}$ define the counting process martingales $M_t^{ij}$ (Equation (\ref{eq:Martingale})) and therefore expected values in the model.
 
Our canonical model is then defined by three technical bases: premium ${\cal B}^P$; valuation ${\cal B}^L$ and accumulation ${\cal B}^A$ (not necessarily the same as ${\cal B}^M$).}

Each pair of technical bases in the model defines {one of} the relationships {\bf R1} to {\bf R3}, set out in Section \ref{sec:Motivation}. Moreover, each relationship so defined is independent of the other, third, technical basis. This set of relationships is well-known in practice (for example, \cite{fisher1965} cited in Section \ref{sec:Motivation}) but there seems to be no coherent account of them in the {technical literature}. We highlight {three} results}.

\begin{bajlist}

\item {(Corollary 1): Each contractual premium $P_i$ can be written as the sum of a pure risk premium and two loadings, one capitalized and taken into surplus at inception, and one taken into surplus only as premiums are paid.}

\item {(Corollary 2): We show that the EPV of total surplus, including initial surplus, is independent of the valuation basis, if the accumulation basis ${\cal B}^A$ is equal to the `true' basis ${\cal B}^M$.}

\item {We define the present value of emerging surplus on accumulation basis ${\cal B}^A$, denoted by $\Gamma_t^{L,A}$ and show how it is related to the present value of surplus $\Gamma_t^L$ defined by \cite{ramlau-hansen1988b}.}

\end{bajlist}

Finally, in the Appendix we discuss possible definitions of retrospective policy value in Markov models, and the usefulness of the concept. {Equality of prospective and retrospective policy values under restricted conditions is a mathematical result in the spirit of a self-financing portfolio. We take the {more practical} requirement to be for a quantity that fairly represents the assets side of the balance sheet, as the prospective policy value does for the liability side, {and our candidate is the accumulation fund or policy account}.}


\acknowledgements

This study is part of the research programme at the Research Centre for Longevity Risk --- a joint initiative of NN Group and the University of Amsterdam, with additional funding from the Dutch Government's Public Private Partnership programme. {We are grateful to Prof Dr Marcus Christiansen for comments on a draft of this paper.} 

\competinginterests

None.

\bigskip
\begin{center}
{\sc Appendix}\\[2mm]
{\sc Retrospective Policy Values}
\end{center}

The main features of definitions of retrospective policy values in a multiple-state model suggested in the literature are as follows.
  
\begin{bajlist}

\item \cite{hoem1969} defined the retrospective policy value as the limit as $N \to \infty$ of the equal share per survivor of the accumulated funds accrued by a cohort of $N$ identical policies; in other words a mathematically rigorous function of actual cashflows. For the simple alive/dead model this gave:

\begin{equation}
{V}_1^-(t) = V_1(t) - V_1(0) / ( v^t \, P_{11}(0,t) ) \label{eq:HoemRetroPolVal}
\end{equation}

\noindent where the `minus' superscript denoted the retrospective policy value. For the general multiple-state model, Hoem defined shares in the collective fund, and {\em chose} parameters such that for $i=1$ equation (\ref{eq:HoemRetroPolVal}) held, and $V^-_i(t) = V_i(t)$ for $i > 1$ .

\item \cite{hoem1988} started with: ``Our purpose is to define $V^-_i(t)$ as a mathematical function for which $V^-_i(t) = V_i(t)$ for all $t$ in $[0,n]$ for as many $i$ as possible''.  There were enough degrees of freedom to assume this equality to be true by {\em fiat} for $i > 1$, leading to equation (\ref{eq:HoemRetroPolVal}) again as the only other constraint.

Hoem also proved conditions for a first-order technical basis (in state 1) to be on the safe-side of a second-order technical basis, along the lines of Lidstone's theorem \citep{norberg1985} but including intuitive conditions on retrospective policy values. He showed that:

\begin{equation}
\bar{V}_1^{*}(t) \le \bar{V}_1(t) = \bar{V}_1^-(t) \le \bar{V}_1^{-*}(t)
\end{equation}

\noindent where `*' denotes the second-order technical basis and the middle equality had previously been shown to hold if $\bar{V}_1(0)=0$.

\item \cite{ramlau-hansen1988a} defined first-order and second-order technical bases, with prospective policy values denoted by $V_i(t)$ and $V_i^0(t)$ respectively $(i \in {\cal S})$ and said: ``let ${}^-V_i(t)$ and ${}^-V_i^0(t)$ denote the retrospective premium reserves [policy values] derived from the two valuation bases.'' Moreover, a condition was imposed (initial state strongly transient, meaning no return to state 1) to ensure that ${}^-V_1^0(0)=0$, and that ${}^-V_i^0(t)$ satisfied Thiele's equation \citep[(3.3)]{ramlau-hansen1988a}. However in the sequel (see Section \ref{sec:SurplusTermMarkov}) the only properties used were: (i) the parameters of the second-order technical basis (to define surplus); and (ii) the two properties of ${}^-V_i^0(t)$ cited above. The retrospective first-order policy value, and prospective second-order policy value, were not used at all. All the results were obtained, in our terms, using a proper valuation basis {$\tilde{\cal B}^L = ({\cal B}^P,P)$} and the accumulation basis ${\cal B}^{M}$.

\item \cite{wolthuis1990} took Hoem's retrospective policy values in (a) above, expressed in matrix form:

\begin{equation}
{\bf V}^-(t) = {\bf V}(t) - v^{-t} \, [P_{11}(0,t)]^{-1} \, (V_1(0),0, \ldots, 0)^T
\end{equation}

\noindent and generalized it to the form:

\begin{equation}
{\bf V}^-(t) = {\bf V}(t) - v^{-t} \, {\bf P}^{-1}(0,t) \, (V_1(0),V_2(0), \ldots, V_m(0))^T,
\end{equation}

\noindent (where ${\bf P}$ is the matrix of occupancy probabilities) also compliant with Thiele's equations (without assuming that state 1 is strongly transient). However, no particular rationale was given, except perhaps additional flexibility.

\item \cite{norberg1991} introduced a very general concept of policy values based on a payment function $A(t)$, specifying the total payments in $[0,t]$, and a discount function $v(t)$, both possibly stochastic. The policy values depended on the following decomposition of the value of $A$ based on payments up to and after time $t$:

\begin{equation}
V(t,A) = - \underbrace{\frac{1}{v(t)} \int_{[0,t]} v(r) \, d(-A)(r)}_{\mbox{\scriptsize def'n} \, \equiv \, V^-(t,A)} + \underbrace{\frac{1}{v(t)} \int_{(t,\infty)} v(r) \, dA(r)}_{\mbox{\scriptsize def'n} \, \equiv \, V^+(t,A)}.\label{eq:ValueDecomp}
\end{equation}

\noindent Then, given a family ${\bf F}$ of sigma-algebras $\{{\cal F}_t\}_{t \ge 0}$, not necessarily a filtration, prospective and retrospective policy values were defined respectively as:

\begin{equation}
V_{\bf F}^+(t,A) = \Mean[ \, V^+(t,A) \mid {\cal F}_t \, ] \qquad \mbox{and} \qquad V_{\bf F}^-(t,A) = \Mean[ \, V^-(t,A) \mid {\cal F}_t \, ]. \label{eq:PVsNorberg} 
\end{equation}

\noindent If ${\cal F}_t$ represents full information about the past then $V_{\bf F}^-(t,A)$ is just the value of actual known cashflows; otherwise a coarser ${\cal F}_t$ represents some grouping of policies defined by missing information. 
The prospective policy value is conventional, and satisfies Thiele's equation, but the retrospective policy value satisfies a different differential equation, generalizing the Kolmogorov forward equations \cite[Section 5E]{norberg1991}. These definitions do not lend themselves to the development of surplus. For all these reasons, we do not pursue these policy values further.

\item \cite{wolthuis1992} {and \cite[Chapter 5]{helwich2007} provide good summaries} of many of the retrospective policy values mentioned above.

\end{bajlist}

Attempts to define retrospective policy values {\em via} a relationship of equality with prospective policy values seem to add little to the analysis of surplus and the real dynamics of a life insurance fund. {Rather, the need is for a quantity that fairly represents the retrospective view of the assets attributed or assigned to, {or accrued by,} a policy or state, just as the prospective policy value represents the need to assign capital to each policy or state. The one may be most influenced by accountancy rules, the other by insurance regulations.} Equality of retrospective and prospective policy values, {on the other hand,} is a mathematical demonstration of circumstances under which, {in expectation only,} the assets acquired under the natural operation of the policies will exactly meet the requirement for capital {--- a `self-financing portfolio' condition}. This is certainly of interest, but we are content {simply to compare unequal supply of and demand for capital. This calls for a model of retrospective accounting}  that is operationally realistic, rather than mathematically ideal, and our definitions of  accumulation basis and policy account have that in mind.

\bigskip

\bigskip

\renewcommand{\section}[1]{\noindent}

\end{document}